\documentclass[12pt]{article}

\input epsf
\usepackage{amssymb,amsmath,wrapfig,subfigure}
\usepackage[matrix,arrow,curve]{xy}
\usepackage{cite}
\usepackage{graphicx,color}
\usepackage[debug,pageanchor=false]{hyperref}
\definecolor{link}{rgb}{.8,.15,.1}
\hypersetup{colorlinks=true,linkcolor=link,citecolor=link,urlcolor=link,linktocpage}

\usepackage{bbold}

\makeatletter
\@addtoreset{equation}{section}
\makeatother

\newcommand{\im}{{\rm Im}}
\newcommand{\re}{{\rm Re}}

\begin{document}

       \begin{titlepage}

       \begin{center}

       \vskip .3in \noindent

       {\Large \bf{Extended Supersymmetry on Curved Spaces}}

       \bigskip

	 Claudius Klare and Alberto Zaffaroni\\

       \bigskip
		 Dipartimento di Fisica, Universit\`a di Milano--Bicocca, I-20126 Milano, Italy\\
       and\\
       INFN, sezione di Milano--Bicocca,
       I-20126 Milano, Italy

       \vskip .5in
       {\bf Abstract }
       \vskip .1in

       \end{center}

       We study $\mathcal{N}=2$ superconformal  theories on Euclidean and Lorentzian four-manifolds
       with a view toward applications to holography and localization. The conditions for supersymmetry
       are equivalent  to a  set of differential constraints including a ``generalised" conformal  Killing spinor equation  depending on various background  fields. 
       We solve these equations  in the general case and give very explicit expressions for the auxiliary fields that we need to turn on to preserve some supersymmetry.
       As opposed to what has been observed for the $\mathcal{N}=1$ case,  the conditions for unbroken supersymmetry turn out to be almost independent of the signature of spacetime, with the exception of few    degenerate cases including the topological twist. 
       Generically, the only geometrical constraint coming from supersymmetry is the existence of a  conformal Killing vector on the manifold, all other constraints determine the background auxiliary fields.

       \noindent

       \vfill
       \eject


       \end{titlepage}
       \section{Introduction}

The study of supersymmetric field theories on curved spaces has led to  many interesting results in recent years, see for example  \cite{Pestun:2007rz,Kapustin:2009kz,Jafferis:2010un,Hama:2010av,Hama:2011ea,Imamura:2011wg,Alday:2013lba}. The approach of coupling the theories to off-shell supergravity,
started in \cite{Festuccia:2011ws},  has led to a classification of supersymmetric backgrounds with Euclidean and Lorentzian signature for ${\cal N} = 1$ theories in four dimensions  \cite{ Klare:2012gn,Dumitrescu:2012ha,Cassani:2012ri,Dumitrescu:2012at} and  for ${\cal N}=2$ theories in three dimensions \cite{Klare:2012gn,Closset:2012ru,Hristov:2013spa}. There are many other related results in various signatures and dimensions  \cite{Blau:2000xg,Jia:2011hw,Samtleben:2012gy,Liu:2012bi,deMedeiros:2012sb,Kehagias:2012fh, Samtleben:2012ua,Kuzenko:2012vd,deMedeiros:2013mca,Cassani:2013dba,Gupta:2012cy}. The analysis is performed by studying the vanishing of the gravitino variation  with  the bosonic  gravity multiplet treated as non-dynamic  background for the field theory.

The cases when the supersymmetric field theories are additionally also conformally invariant can be studied by coupling the theory to conformal supergravity or, alternatively, using holography \cite{Klare:2012gn}. For theories with four supercharges, it follows that a SCFT retains some supercharges on a curved space only if it admits a solution to the conformal Killing spinor (CKS) equation  \cite{Klare:2012gn,Cassani:2012ri}. 
The vanishing of the gravitino equation is indeed equivalent to a charged version of the CKS equation. For theories with higher supersymmetry, the analysis is more involved since the conformal gravity multiplet contains another dynamical fermion in addition to the gravitino. Its supersymmetry variation leads to other differential equations which should be added to a generalised CKS equation involving various background fields.

In this paper we analyse the ${\cal N}=2$ case in four dimensions, both in Lorentzian and Euclidean signature,  and we determine the general couplings to auxiliary backgrounds fields that preserve some of the extended supersymmetry.  The  Lorentzian case where part of the superconformal invariance is gauge fixed by compensators  has already been analysed and completely solved in \cite{Gupta:2012cy} for applications to black hole entropy\footnote{
We thank Sameer Murthy for pointing out to us their results, which have substantial overlap with section 2.}. Here we generalise the result to the full CKS equation and extend it to the Euclidean case. Although some of the supersymmetry constraints are differential, we show that they can always be solved by choosing appropriate local coordinates.   We give very explicit expressions for the auxiliary fields that we need to turn on to preserve some supersymmetry. In general, the auxiliary fields are not unique and there is some arbitrariness in their choice.  
 The general result  turns out to be almost independent of the signature of spacetime, with the exception of  few degenerate cases.

A word of caution should be spend for Euclidean theories.  Of course, it is well known that we can define a consistent ${\cal N}=2$ theory on {\it any} Euclidean  four 
manifold by a topological twist \cite{Witten:1988ze}. 
The twist was analysed in the language of this paper long time ago \cite{Karlhede:1988ax}. However this is not the only way of preserving supersymmetry. 
For example, we can  define a SCFT on any conformally flat curved space just by a conformal mapping from flat space. This is the case of the theories studied in \cite{Pestun:2007rz}, for example. In between these two extreme situations there is a full spectrum of possibilities with different auxiliary fields that is investigated in this paper.

In Lorentzian signature  the condition for preserving some supersymmetry for an ${\cal N}=2$ theory in four dimensions is equivalent to
the existence of a conformal Killing vector (CKV).   Of course, it is simple to see that the existence of a (charged)  CKS implies the existence of a CKV.
The converse is also true: the geometric constraints following from supersymmetry just amount to the existence of a CKV, all other constraints determine
the background auxiliary fields of the gravity multiplet.  Analogously to three dimensions \cite{Hristov:2013spa}, the CKV can be null or time-like, the null case being related to the existence of an ${\cal N}=1$ subalgebra,  a case discussed in detail in \cite{Cassani:2012ri}.

In Euclidean signature, the conditions for supersymmetry involve two symplectic Majorana Weyl spinors of opposite chirality. There is a degenerate case where we preserve supersymmetry by using spinors
of one chirality. The conditions of supersymmetry collapse to a non-abelian version of the CKS equation.  The topological twist  \cite{Witten:1988ze} falls in this class;  the spinor is made covariantly constant by  identifying the $SU(2)$ R-symmetry of the theory with the spin connection of the four manifold. This works for any four-manifold. In this paper we analyse in detail the general case where supersymmetry is preserved using
spinors of both chirality. In this case, as in Lorentzian signature, the condition for preserving some supersymmetry  is equivalent to
the existence of a conformal Killing vector (CKV),  all other constraints determining the background auxiliary fields for which we provide general expressions.

Our results are similar to the Lorentzian ones for theories with four supercharges in three and four dimensions \cite{Cassani:2012ri,Hristov:2013spa} and somehow different from the Euclidean results  with four supercharges where the geometric  constraints require the manifold to be complex in four dimensions  \cite{ Klare:2012gn,Dumitrescu:2012ha} and to possess a suitably constrained contact structure in three
\cite{Klare:2012gn,Closset:2012ru}. 

In general,  Poincair\'e supergravities arise from    conformal supergravity through the coupling to compensator multiplets which gauge fix the redundant symmetry. Some of our results can be used   also to define general supersymmetric field theory on curved space, although we do not discuss this issue in our paper. 

The paper is organized as follows. In section \ref{sec:lor} we discuss the Lorentzian case. We first  review the field content of ${\cal N}=2$ conformal supergravity and we write
the supersymmetry variations which include a {\it generalised} conformal Killing spinor equation. We then discuss the geometrical structure induced by a pair of chiral spinors and the technical tools
we will be using in this paper. Section \ref{sec:L:ClassGeo} contains the main results for the Lorentzian case: the proof that any manifold with a conformal Killing vector supports some supersymmetry
and the explicit expressions for the background auxiliary fields.  In section \ref{sec:euclid} we discuss the Euclidean  case in close parallel with the Lorentzian one, the main results being
presented in section \ref{sec:E:ClassGeo}. Explicit examples, ranging from various topological twists to the supersymmetry on round and squashed spheres are then discussed. We
finish with some comments.

       \section{The Lorentzian Case}\label{sec:lor}
       In this section we will identify Lorentzian four-dimensional manifolds that preserve some of the supersymmetries of an $ \mathcal{N}=2 $ conformal field    theory.   It turns out that the only geometrical constraint to be imposed on the manifold is the existence of a null or timelike conformal Killing vector, thus
       generalising the results obtained in \cite{Cassani:2012ri} for the ${\cal N}=1$ case. Compared with ${\cal N}=1$ supersymmetry, the CKV is now
       allowed to be time-like. The situation is very similar to the three-dimensional case discussed in \cite{Hristov:2013spa}. 
             
             Several of the results in this section have been already obtained in \cite{Gupta:2012cy} in the context of the study of black hole entropy. In particular, the case where the spinor $\eta$  associated with the   conformal supersymmetry vanishes has been  analysed in detail in that reference and  the general background fields allowing for some supersymmetry have been found.  When $\eta=0$, the conformal Killing vector actually becomes Killing. 
                           
       Our strategy is to couple the theory  to some background supergravity and then to freeze the 
       fields in the gravity multiplet to rigid background values \cite{Festuccia:2011ws}. 
       The matter couplings to supergravity in this rigid limit become the curvature couplings of the field theory in curved space.
       If the field theory is conformally invariant, it is natural to couple it to conformal supergravity. 
       This can be understood also from a different perspective, as we can study a CFT  via its holographic dual.
       Typically, minimal  supergravity in asymptotically locally $AdS$ spaces reduces to conformal supergravity on the (non-trivial) boundary, see \emph{e.g.\ }\cite{Balasubramanian:2000pq}.
       This has been discussed recently in \cite{Klare:2012gn, Cassani:2012ri} and seems to be true also for situations with more supersymmetry.
       In fact, expanding Roman's $\mathcal{N}=4$ $5$d gauged supergravity \cite{Romans:1985ps} in an asymptotically AdS space, 
       one finds \cite{Ohl:2010au} the conditions of $\mathcal{N}=2$ superconformal supergravity in four dimensions 
       \cite{deWit:1979ug, deWit:1980tn} on the boundary.
       A similar analysis is true for general boundaries, where the bulk is only \emph{locally} asymptotically AdS.

\subsection{The multiplet of conformal supergravity}   \label{sec:conf}    
       The structure of conformal supergravity in $4$ dimensions is nicely reviewed in  \cite{LectParis}\footnote{
       In our discussion of conformal supergravity in $4$ dimensions, we mostly follow the conventions of \cite{LectParis} with few changes that are discussed in the Appendix.
       We have also redefined the gauge field and the scalar.
       }.
       The independent field content of the gravity (Weyl) multiplet is
       \begin{align}
	 & g_{\mu \nu} \,  && \psi_\mu^i \,  && T^+_{\mu \nu} \, & & \tilde{d} \,  && \chi^i \, && A_{\mu 0} \,  && A_\mu{}^i{}_j \,
	 \label{eq:Weylmplt}
       \end{align}
       where $A_{\mu 0}$ and   $A_\mu{}^i{}_j = A_{\mu x} \sigma^x \,^i{}_j$ are the gauge fields of the $U(1)$ and $SU(2)$ R-symmetry, respectively
       (with $\vec{\sigma}$ the usual Pauli matrices),
       $T^+$ is a (complex) self-dual tensor, $\tilde d$ a scalar field and $\chi^i$ the  dilatino.
       The fermions are chiral spinors in the $\bf{2}$ of $U(2)$.
       The fermionic part of the supersymmetry variations is
       \begin{equation}  \label{eq:L:BoundSusy}
	 \begin{aligned}
	 \delta \psi_\mu^i &= \nabla^A_\mu \epsilon^i_+ + \frac{1}{4} T^+_{\mu \nu} \gamma^\nu \epsilon_-^i - \gamma_\mu \eta_-^i \, 
	   \\
	 \delta \chi^i &= 
	 \frac{1}{6} \nabla^A_\mu T^+ \,^\mu{}_\nu \gamma^\nu \epsilon^i_- 
	 -\frac{i}{3} R^{SU(2)}\,^i{}_j \cdot \epsilon_+^j 
	 +\frac{2i}{3} R^{U(1)} \cdot \epsilon_+^i
	 + \frac{\tilde{d}}{2} \epsilon_+^i
	 +\frac{1}{12} T^+_{\mu \nu} \gamma^{\mu \nu} \eta_+^i
	 \end{aligned}
       \end{equation}
with
\begin{align*}
  \nabla^A_\mu \epsilon_+^i &= \nabla_\mu \epsilon_+^i - i A_{\mu \alpha} \sigma^\alpha{}^i{}_j \epsilon_+^j \, \\
  \nabla^A_\mu T^+ &= \left( \nabla_\mu -2i A_{\mu 0} \right) T^+ \\
  R^{SU(2)} \,^i{}_j &=  (\partial_\mu A_{\nu x} + A_{\mu y} A_{\nu z} \epsilon^{yz}{}_x) \sigma^x \,^i{}_j \gamma^{\mu \nu} \, \\
  R^{U(1)} &= \partial_\mu A_{\nu 0} \gamma^{\mu \nu} \, 
\end{align*}
where we have introduced 
       \begin{align}
	 \sigma^\alpha \,^i{}_j = (\mathbb{1}, \vec{\sigma}){}^i{}_j  \, .
	 \label{eq:L:sigma}
       \end{align}
       The spinor doublets $\epsilon^i$ and $\eta^i$ are the $Q$- and $S$- supersymmetry parameters, respectively.
Our conventions  are summarized in appendix \ref{sec:convent}, in particular $\alpha,\beta,\dots = 0,1,2,3$ and $x,y,\dots = 1,2,3$. 


To  preserve some supersymmetry on a  manifold with metric $g_{\mu \nu}$ we need to find a configuration
of auxiliary fields of  the Weyl multiplet that solves $\eqref{eq:L:BoundSusy}$. Obviously, this is not always possible and we now 
want to analyse in which cases it can be done.   The special case where $\eta^i=0$ has already been analysed in \cite{Gupta:2012cy}, in the following  we extend this result to the general case.

We can eliminate $\eta^i$ by taking the $\gamma$-trace of the first equation
$\eta_-^i = \frac14 D^A \epsilon_+^i$, with $D^A = \gamma^\mu \nabla^A_\mu$. 
Note that $T$ drops out of the computation since $\gamma_\mu T^+ \gamma^\mu = 0$.
The supersymmetry condition can then be re-written as\footnote{
See also \cite{Hama:2012bg} for a similar presentation.
}
   \begin{subequations} \label{eq:L:SUSY}
       \begin{align}
          \nabla^A_\mu \epsilon^i_+ + \frac{1}{4} T^+_{\mu \nu} \gamma^\nu \epsilon_-^i - \frac14 \gamma_\mu D^A \epsilon_+^i &= 0 
	  \label{eq:L:SUSY:gravitino} \\
	 \nabla^A_\mu T^+\,^\mu\,_\nu \gamma^\nu \epsilon_-^i + D^A D^A \epsilon_+^i 
	 + 4i \nabla_\mu A_{\nu 0} \gamma^{\mu \nu} \epsilon_+^i +2d\,\epsilon_+^i &=0
	 \label{eq:L:SUSY:dilatino}
       \end{align}
   \end{subequations}
     where we have used the first equation and we have redefined $d\equiv \tilde{d} + \frac16 R$, with $R$ being the curvature scalar.

     The gravitino equation $\eqref{eq:L:SUSY:gravitino}$ can be seen as a generalisation of the 
     charged conformal Killing spinor equation found in the $\mathcal{N} = 1$ case \cite{Klare:2012gn}.
     For $T^+= 0$ we simply obtain a non-abelian version of the CKS equation. In general,
     the situation is more involved due to the presence of the  tensor $T^+$, however  the  equation $\eqref{eq:L:SUSY:gravitino}$ 
     shares many similarities with the  CKS equation,  in particular, it is  conformally covariant. 
     If the doublet $\epsilon_+^i$ is a solution to the equation with metric $g_{\mu \nu}$,       the rescaled doublet $e^{\lambda/2} \epsilon_+^i$ is a solution to the equation with rescaled metric $e^{2\lambda} g_{\mu \nu}$.
     In particular, the tensor has conformal weight $+1$
     \[
     T^+_{\mu \nu} \rightarrow e^{\lambda} T^+_{\mu \nu}\, .
     \]
     
     A further complication that affects all extended supergravities  is the presence of the dilatino equation.  As a difference with the generalised CKS equation
     $\eqref{eq:L:SUSY:gravitino}$ which can be analysed in terms of a set of algebraic constraints for the geometric quantities involved, as in  \cite{Klare:2012gn, Cassani:2012ri}, the dilatino equation contains derivatives of the auxiliary fields and it seems to be more complicated to analyse. However, we will show how to extract the relevant information from it. It turns out that it is only the gravitino equation that restricts the geometry of the space-time, 
     while the conditions coming from the dilatino equation $\eqref{eq:L:SUSY:dilatino}$ merely fix some of the background field values.

    We stress that the supersymmetry variations of conformal supergravity do not depend on the explicit matter content of the  field theory.
    Our result will therefore be valid for any conformal theory with rigid supersymmetry. We also notice that all Poincair\'e supergravities arise from
    conformal supergravity through the coupling to compensator multiplets which gauge fix the redundant symmetry. Some of our results can also be used 
    to define general supersymmetric field theories on curved space, although we do not discuss that issue further in this paper.  The close relation
    between ${\cal N}=1$ CKS and new minimal supergravity spinors is described in detail in \cite{ Klare:2012gn, Cassani:2012ri}.  For the ${\cal N}=2$
    case we just observe that we can always set $\eta^i=0$  after a partial gauge fixing of the superconformal symmetry using a hypermultiplet compensator.
    The results in \cite{Gupta:2012cy}, where this gauge fixing is done, can therefore be used to  define general supersymmetric field theories on curved   space. As can be seen in \cite{Gupta:2012cy}, or by specialising the results in section \ref{sec:L:ClassGeo} to the case $\eta^i=0$, the conditions of supersymmetry now requires that the conformal Killing vector is actually Killing. This is similar to what happens in the ${\cal N}=1$ case \cite{Cassani:2012ri}.

     The supersymmetry transformations and the most general Lagrangian for matter couplings to conformal supergravity can be found in 
     \cite{LectParis}\footnote{An analysis of  supersymmetry Lagrangians using conformal supergravity has also appeared in \cite{deWit:2011gk} where various explicit solutions  have been discussed.}. 
     For completeness, here we give the Lagrangian for the vector multiplet $(\phi, W_\mu, \psi_+{}_i, Y^{ij})$,
     \begin{multline} \label{eq:L:Lagrangian} 
       \mathcal{L}^{\text{Lorentz}} = 
	 d \, \phi \bar \phi + \nabla^A_\mu \phi \nabla^A{}^\mu \bar \phi + \frac{1}{8} Y^{ij} Y_{ij} - g \left[ \bar \phi , \phi \right]^2
	 +\frac{1}{8} F_{\mu \nu} F^{\mu \nu} \\
	  +\big\{ 
	 \frac{1}{4} \bar \psi_-^i D^A \psi_+{}_i + \frac{1}{2}g \bar \psi_-^i \left[ \phi, \psi_-{}_i \right] 
	 -\frac14 \phi F_{\mu \nu} T^+{}^{\mu \nu}
	 -\frac{1}{16} \phi ^2 T^+_{\mu \nu} T^+{}^{\mu \nu} 
	 + \text{h.c.} \big\} 
     \end{multline}
     where $Y_{ij}= \left( Y^{ij} \right)^\ast = \epsilon_{ik} \epsilon_{jl} Y^{kl}$ is a triplet of real $SU(2)$ scalars.

     In the next sections we will classify the geometries in which one can solve $\eqref{eq:L:SUSY}$.
     The background fields that we will determine in terms of the geometry can be coupled to 
     arbitrary vector and hypermultiplets as in $\eqref{eq:L:Lagrangian}$, giving rise to supersymmetric field  theories on curved space.

\subsection{The geometry of spinors}\label{sec:geometryspinors}     

     In our analysis we follow a similar formalism as in \cite{Klare:2012gn, Cassani:2012ri, Hristov:2013spa,Gupta:2012cy} .
       We want to analyse the geometry that is defined by a chiral spinor doublet of $U(2)$.
       To this end, let us look at the two spinor bilinears
       \begin{equation} \label{eq:L:sz} 
	 \begin{aligned}
	   s &= \frac12 \epsilon_{ij} \, \bar \epsilon_-^i \epsilon_+^j \\
	   z_\mu &= \frac12 \bar \epsilon_+{}_i \gamma_\mu \epsilon_+^i
	 \end{aligned}
       \end{equation}
       with $z_\mu z^\mu = - \| s \|^2$. It is a simple consequence of the generalised CKS equation that $z$ is a conformal Killing vector.
       To proceed, we need to distinguish two situations.

       When $s=0$, the two spinors in the doublet are linearly proportional $\epsilon_+^1 \propto \epsilon_+^2$ and $z_\mu$ is null.
       We essentially fall in the  $\mathcal{N}=1$ case that has been studied in \cite{Cassani:2012ri}. In fact, by a gauge transformation we can
       always set one of the spinors to zero. By further restricting the gauge fields to a suitable abelian subgroup and by setting the tensor $T^+$
       to zero, we obtain the charge CKS equation discussed and solved in \cite{Cassani:2012ri}. In that reference it is shown that any manifold
       with a null CKV supports supersymmetry. It is enough to turn on  a background abelian gauge field whose explicit form is given in \cite{Cassani:2012ri}\footnote{It is not excluded that {\it more} general solutions with non abelian gauge fields and a non-vanishing tensor exist but
        we will not discuss this case in our paper.}. 
       
       We therefore refer to  \cite{Cassani:2012ri} for the case with $s=0$ and from now on we will discuss the non-degenerate situation where $s\equiv e^{B+i\beta} \neq 0$.
       Apart from the complex scalar $s$, 
       a generic chiral doublet of spinors in four dimensions defines four real vectors $e^n_\mu$ and three self-dual two-forms $\eta^x_{\mu \nu}$
       \begin{equation} \label{eq:L:bilinears} 
	 \begin{aligned}
	   e^B  \sum_\alpha e^\alpha_\mu \, \sigma^\alpha{}^i{}_j &= \bar \epsilon_+{}_j \gamma_\mu \epsilon_+^i \\ 
	   2 e^{B+i\beta} \eta^x_{\mu \nu} \, \sigma^x{}^{ij} &= \bar \epsilon_-^i \gamma_{\mu \nu} \epsilon_+^j
	 \end{aligned}
       \end{equation}
       where $\sigma^\alpha{}^{ij} = \sigma^{\alpha i}{}_k \epsilon^{kj}$ and $e^0_\mu = e^{-B} z_\mu$.
       The self-duality condition  reads
       \[
       \epsilon_{\mu \nu}{}^{\rho \sigma} \eta^x_{\rho \sigma} = 2 i \eta^x_{\mu \nu} \, .
       \]
       
       Using the Fierz identities one can show that $e^\alpha_\mu$ is a tetrad, $e^\alpha \cdot e^\beta = \eta^{\alpha\beta}$. 
       One can also show that in this frame the $\eta_x$'s have components
       \begin{align} \label{eq:L:etaDef}
	 \eta_x^{\alpha\beta} = e^\alpha_\mu e^\beta_\nu   \eta_x^{\mu \nu} &= \delta_0^\alpha \delta_x^\beta -\delta_0^\beta \delta_x^\alpha -i \epsilon_x{}^{\alpha\beta} \,.
       \end{align}
       We give some details in the appendix.

        Note that the information contained in the spinor doublet can also be written in bispinor language\footnote{See \cite{Tomasiello:2011eb}, in particular Appendix A, for a nice  review of the formalism.}
       \begin{equation}
	 \label{eq:L:bispinors}
       \begin{aligned}
	 \epsilon_+^i \bar \epsilon_{+j} &= \frac{e^B}{4} \sum_\alpha \left( e^\alpha + i \ast e^\alpha \right) \sigma^{\alpha}{}^i{}_j  \\
	  \epsilon_+^i \epsilon_-^j &= \frac{e^{B+i\beta}}{4} \eta_\alpha \sigma^\alpha{}^{ij} 
       \end{aligned}
       \end{equation}
       where $\eta_0 = -\left( 1+\gamma \right)$, $\eta_x = \frac12 \eta_x{}_{\mu \nu} \gamma^{\mu \nu}$ 
       and $e^{\alpha} = e^{\alpha}_\mu \gamma^\mu$. Here, $i \ast e^\alpha = \gamma e^\alpha $.

       From the definition in $\eqref{eq:L:bilinears}$ we see that $e^x$ and $\eta^x$ transform as 
       vectors under the action of  the $SU(2)$ R-symmetry. From $\eqref{eq:L:bilinears}$ it also follows that we can gauge away $\beta$ by a $U(1)$  R-symmetry    transformation.
       Similarly, since our equations are conformally covariant, we can -- at least locally -- 
       also set $B$ to zero by an appropriate Weyl rescaling.
       
       We will find useful to work in the frame defined by $e^n$ where the action of the gamma matrices on the spinors takes a very simple form.
      For example it is easy to show that
       \begin{equation} \label{eq:L:gammaxeps}
       \begin{aligned}
	 \gamma^\alpha \epsilon_-^i &= e^{-i\beta} \sigma^\alpha{}^i{}_j \epsilon_+^j \\
	 \gamma^{\alpha\beta} \epsilon_+^i &= \eta_x^{\alpha\beta} \sigma^x{}^i{}_j \epsilon_+^j \, .
       \end{aligned}
     \end{equation}
       From $\eqref{eq:L:gammaxeps}$ it readily follows that, in the frame defined by $e^n$, the spinors are constant, up to an overall norm factor. 

       Before we attack the supersymmetry conditions, 
       note that it is useful to choose a covariant basis for the space of chiral spinor doublets\footnote{ \label{fn:bases}
       Another obvious choice of basis would be $\gamma_\mu \epsilon_-^i$. The two bases are related by $\eqref{eq:L:gammaxeps}$.}
       \begin{equation}
	 \sigma^\alpha{}^i{}_j \epsilon_+^j \, .
	 \label{eq:L:SpinorBasis}
       \end{equation}
       We can expand in particular the covariant derivative of a spinor in this basis
       \begin{equation}
	 \nabla_\mu \epsilon_+^i = P_{\mu \alpha} \sigma^{\alpha i}{}_j \epsilon_+^j
	 \label{eq:L:DefP}
       \end{equation}
       where we call the coefficients $P$ \emph{intrinsic torsions}.
       More explicitly, in our frame with constant spinors one has 
       \begin{align}
	 & P_{\mu x} = \frac14 \omega_\mu{}^{\alpha \beta} \eta_x{}_{\alpha \beta} && P_{\mu 0} = \frac12 \partial_\mu (B +i\beta)
	 \label{eq:L:Torsions}
       \end{align}
       where $\eta_x{}_{\alpha \beta}$ projects on the self-dual part of the spin connection.
       A similar parametrisation has appeared in \cite{Rosa:2013jja}.

       \subsection{Solving the supersymmetry conditions}
       \label{sec:L:ClassGeo}
        In this section we show that all manifolds with a timelike conformal Killing vector (CKV) admit  a solution to the 
       equations $\eqref{eq:L:SUSY}$.

       Let us analyse the two conditions for unbroken supersymmetry $\eqref{eq:L:SUSY}$ separately. 
       We define the symmetric traceless part of the torsion\footnote{It will be convenient 
       to 
       consider $P_{(\alpha\beta)} $ and $A_{(\alpha\beta)}$ as four-by four matrices   where the indices are raised   and lowered with $\eta_{\alpha\beta}$. All our formulae will be valid in the frame define by the $e^\alpha$.}
       \begin{align*}
	 p_{\alpha\beta} & \equiv P_{(\alpha\beta)} - \frac{\eta_{\alpha\beta}}{4} P_\gamma^\gamma 
       \end{align*}
       and the gravitino equation $\eqref{eq:L:SUSY:gravitino}$ is readily solved by requiring 
       \begin{subequations}
	 \label{eq:L:solutionGravitino}
         \begin{align}
	    \re (p_{\alpha\beta})&= 0  \label{eq:L:CKSp}\\
	    \im \left( p_{\alpha\beta} \right) &= A_{(\alpha\beta)} - \frac{\eta_{\alpha\beta}}{4} A_\gamma^\gamma  \label{eq:L:ASTrless}\\
	    T^+_{\alpha\beta} &= -4 e^{i\beta} \left( P^+_{\alpha\beta} - i A^+_{\alpha\beta} \right) \label{eq:L:Tplus} \, .
         \end{align}
       \end{subequations}
     The first line is a constraint on the geometry of the manifold, 
     while the second and the third line are merely fixing some of the background fields in terms of this geometry. 
     Let us discuss this more explicitly.

     The geometrical interpretation of $\eqref{eq:L:CKSp}$ is very simple, it is equivalent to $z =e^B e^0$ being a conformal Killing vector, 
     i.e.\ fulfilling
     \begin{align}
         \nabla_{(\alpha}  z_{\beta)} = \lambda \eta_{\alpha\beta} \, .
       \label{eq:L:CKVdef}
     \end{align}
     To see this, note that $\eqref{eq:L:CKVdef}$ written in our frame reads
     \begin{align}
       &\lambda = -e^{B} \partial_0 B 
       &
       & \omega_{(x}{}^{0}{}_{y)} = \delta_{xy} \partial_0 B &
       & \omega_0{}^{0}{}_x = \partial_x B
       \label{eq:L:CKVomega}
     \end{align}
     where we defined $\partial_n B \equiv e^\mu_n \partial_\mu B$ etc.
     Then, using $\eqref{eq:L:Torsions}$, it is easy to see that this is equivalent to $\re \left( p_{\alpha \beta} \right)=0$. 

     The other two equations determine the ``symmetric traceless" part of the gauge field and the value of the tensor field, respectively.
     In our frame, $\eqref{eq:L:ASTrless}$ reads 
     \begin{align}
       A_{(\alpha\beta)} - \frac{\eta_{\alpha\beta}}{4} A_\gamma^\gamma &= 
       -\frac14 \Big( \epsilon_{uv(\alpha} \omega_{\beta)}{}^{uv} +\frac12 \delta_{(\alpha}^{\;0} \partial_{\beta)} \beta
       - \frac{\eta_{\alpha\beta}}{4} (\epsilon_{yz}{}^x \omega_x{}^{yz} +\frac12 \partial_0 \beta)  \Big)
       \label{eq:L:ASTrlessFrame}
     \end{align}
     while $\eqref{eq:L:Tplus}$ fixes $T^+$ in terms of the ``antisymmetric" part of the gauge field, which we will determine in the next paragraph.
     
     It actually turns out that $z$ being a CKV is the only geometrical constraint for unbroken supersymmetry. 
     The dilatino equation $\eqref{eq:L:SUSY:dilatino}$ gives no extra conditions for the manifold,
     in fact it has exactly the right amount of degrees of freedom to fix the background fields which are yet undetermined.
     There are $8$ components and we still have to determine the values of $A_{[\alpha \beta]}$, $A_\gamma^\gamma$ and $d$.

     To this end it is useful to parametrise the gauge fields satisfying  \eqref{eq:L:ASTrlessFrame} as
     \begin{equation} \label{eq:L:solA}
      \begin{aligned}
       A_{x0} &\equiv -b_x +\frac12 \partial_x \beta \\
       A_{0x} &\equiv b_x - \frac14 \epsilon_{xuv} \omega_0{}^{uv} \\
       A_{00} &\equiv -\frac14 \alpha +\frac12 \partial_0 \beta\\
       A_{xy} &\equiv \epsilon_{xy}{}^z a_z + \frac14 \delta_{xy} \alpha - \frac14 \epsilon_{uvy} \omega_{x}{}^{uv} \, .
      \end{aligned}
     \end{equation}
     The new quantities $\alpha, a_x$ and  $b_x$ parametrise the ``trace" and the ``antisymmetric" part of the gauge field $A_{\alpha\beta}$ \footnote{Note that  $ a_x, b_x$ correspond precisely to the imaginary antisymmetric part of the (``twisted") intrinsic torsions  
     \begin{equation} \label{eq:L:TorsionsInAlpha} 
       \begin{aligned}
	 P^A_{x0} &= - P^A_{0x} =  i b_x +\frac12 \partial_x B \\
	 P^A_{xy} &= -i \big( \epsilon_{xy}{}^za_z +\frac{1}{4} \delta_{xy} \alpha \big) -\frac12 \omega_x{}^0{}_y
       \end{aligned}
     \end{equation}
     where the superscript $A$ denotes twisting with the $U(2)$ gauge field  $P_{\alpha \beta}^A = P_{\alpha \beta } - i A_{\alpha \beta}$.}.
     Then, the background value for the complex tensor $T^+$ becomes
     \begin{align} \label{eq:L:TplusShifted} 
       \frac12 e^{-i\beta} T^+{}_{0x} = i(b_x - \frac14 \epsilon_{x}{}^{yz} \omega_y{}^{0}{}_z) + (a_x +\frac12 \partial_x B)\, .
     \end{align} 
     The spatial components are given via self-duality $T^+_{xy} = i \epsilon_{xy}{}^z T^+_{0z}$.

     In this language the dilatino equation can be re-written in a particularly simple form. 
     This re-writing is a bit lengthy but straightforward. As a result we find eight real equations, coming from the real and immaginary
     part of $\eqref{eq:L:SUSY:dilatino}$ in the basis $\eqref{eq:L:SpinorBasis}$
     \begin{equation} \label{eq:L:DilatinoResult} 
       \begin{aligned}
	 \partial_0 a_x + ( \omega_x{}^{y}{}_0 - \omega_0{}^{y}{}_x ) a_y &= -\frac12 e^{-B} \partial_x (e^B \partial_0 B) \\
	 \partial_0 b_x + (\omega_x{}^y{}_0 - \omega_0{}^y{}_x) b_y &=0 \\
	 \partial_0 (e^B \alpha) &= 0 
       \end{aligned}
     \end{equation}
     \begin{equation} \label{eq:L:DilatinoResult:4} 
       \begin{aligned}
	 d = 2 (\partial_x + \omega_u{}^u{}_x + a_x) a^x - (\epsilon_x{}^{yz} \omega_y{}^0{}_z &+ 2 b_x) b^x + \frac14 \alpha^2\\
	     &+\frac14 (\omega_x{}^{0y})^2 
	    +\partial_0 \partial_0 B -\frac{1}{2} (\partial_x B)^2  +\frac{5}{4} (\partial_0 B)^2
       \end{aligned}
     \end{equation}
     where expressions like $\partial_0 \partial_0 B$ are to be understood as $e_0^\mu \partial_\mu (e_0^\nu \partial_\nu B)$ etc.
     The seven equations $\eqref{eq:L:DilatinoResult}$ determine the missing parts of the gauge field in terms of the geometry, 
     while equation $\eqref{eq:L:DilatinoResult:4}$ fixes the scalar $d$.
   
     Note that we can solve these equations, at least locally, by choosing a particular set of coordinates. 
     So far, $\eqref{eq:L:DilatinoResult}$ is valid for any frame in which $z$ is conformal Killing. 
     We can choose a Weyl representative of the metric such that it becomes \emph{Killing} instead. 
     Then, one can choose coordinates such that $z=\partial/\partial t$ and the metric can locally be written as 
     \begin{align} \label{eq:L:MetricKilling} 
       ds^2 = -e^{2B} (dt + 2\mathcal{F})^2 + \mathcal{H}_{ij} dx^i dx^j
     \end{align} 
     where $B$, $\mathcal{F}$ and $\mathcal{H}$ do not depend on $t$. 
     $\mathcal{F}$ is a one-form on the spatial part transverse to $z$.
     As a one-form, we have 
     \begin{equation} \label{eq:L:zoneform} 
       z= e^{2B}(dt+2\mathcal{F}) \, .
     \end{equation} 
     In such coordinates we have additional symmetries of the spin connection. 
     If we also choose a $t$-independent frame for the spatial dimensions, we have $e^0 \cdot de^x = 0$, which implies 
     \begin{align} \label{eq:L:symmofspinconnection} 
       \omega_0{}^x{}_\alpha + \omega_\alpha{}^{x0} = 0
     \end{align} 
     and we gain more explicit expressions for the spin connection
     \begin{align} \label{eq:L:SpinConnCoord} 
       \omega^0{}_x = \partial_x B \; e^0 + e^B (d \mathcal{F}_{yx} ) \; e^y &
       & \omega^x{}_y = \tilde \omega^x{}_y - e^B ( d \mathcal{F})_{xy} \, e^0
     \end{align} 
     where $\tilde \omega^{xy} = \omega_z{}^{xy} e^z$ is the spin connection on the three-dimensional space transverse to $z$.
     
     Taking into account the symmetry $\eqref{eq:L:symmofspinconnection}$,
     the differential constraints $\eqref{eq:L:DilatinoResult}$ boil down to
     \begin{align} \label{eq:L:DilatinoResultCoord} 
       \partial_t \alpha = \partial_t a_x = \partial_t b_x &= 0  \, .
     \end{align} 
     We see that we have the freedom to choose arbitrary values for $a_x, b_x$ and $\alpha$ as long as they do not depend
     on the isometry coordinate  $t$.
     One simple solution can be obtained for example by requiring $T^+$ to vanish, yielding
     \begin{equation}\label{eq:lortwist} 
       \begin{aligned}
	 T^+_{\alpha \beta} &= 0 \\
	 A_{00} &= -\frac{\alpha}{4} \\
	 A_{x0} &= \frac{1}{2} \big( \partial_x \beta -e^B (\tilde \ast \, d \mathcal{F})_x \big) \\
         A_{0x} &=  e^B (\tilde \ast \, d \mathcal{F})_x \\ 
	 A_{xy} &= -\frac14 \big( \epsilon_{uvy}  \omega_x{}^{uv} + 2 \epsilon_{xy}{}^z \partial_z B  - \delta_{xy} \alpha \big)\\
     \end{aligned}
     \end{equation}
     where $\tilde \ast$ is the three-dimensional Hodge dual, acting on forms living on the spatial part.
     The value for the scalar field follows immediately from $\eqref{eq:L:DilatinoResult:4}$.

We can explicitly check that, by restricting to the case where $\eta^i=0$, we reproduce the results found in \cite{Gupta:2012cy}. In this particular case
the conformal Killing vector becomes Killing.

     To summarise, we can preserve some 
   supersymmetry on any manifold with a time-like Killing vector. 
     To this end, we have to turn on the ``symmetric traceless" part of the background gauge field as determined in $\eqref{eq:L:ASTrlessFrame}$,
     the background tensor field as in $\eqref{eq:L:TplusShifted}$, and the background scalar as in $\eqref{eq:L:DilatinoResult:4}$.
     Upon picking special coordinates $\eqref{eq:L:MetricKilling}$, we are free to choose a $t$-independent  ``trace" and ``antisymmetric" part of the gauge field.  

     \subsection{Examples, comments and possible extensions to higher dimensions}\label{sec:higher}

    We have seen that any metric with a non-vanishing timelike Killing vector (and all conformally equivalent metrics) supports some supersymmetry. The general form of    such metrics, up to Weyl rescaling, is given in \eqref{eq:L:MetricKilling}.  
    

     In the particular case where the manifold  is the direct product $\mathbb{R}\times M_3$ with $M_3$ being an Euclidean three-manifold
     we can just use the $SU(2)$ gauge group and set $A_{\mu0}=0$ and $T^+=0$.    
    This is obvious from our solution 
     $\eqref{eq:lortwist}$, which for a direct product space collapses to 
     $A_{\mu x}=-\frac14 \epsilon_{uvx} \tilde \omega_\mu^{uv}$,
     with all other background fields vanishing. We have taken the spinor to be constant.
     The $SU(2)$ R-symmetry background field is identified with  the $SO(3)$ spin connection on $M_3$, 
     making the spinor covariantly constant.
     This is a
    particular instance of the Euclidean Witten twist applied to the three-manifold $M_3$. These kind of solutions have an interesting application  in holography as boundary theories of supersymmetric
    non-abelian black holes in AdS$_5$ \cite{kirilrota}.

    We should mention that we have assumed up to now that  the norm of the Killing vector were nowhere vanishing. In the cases where 
    it becomes null on some sub-manifold more attention should be paid to the global properties of the solution. Examples of this kind are
    discussed explicitly in the three-dimensional case in \cite{ Hristov:2013spa}.  
    
     We can also make some speculations about extended supersymmetry in higher  dimensions. 
     Curiously, a counting of degrees of freedom in the Weyl multiplet of conformal supergravity
     suggests the possibility that theories with $8$ supercharges generally preserve some of their supersymmetry 
     precisely on manifolds with a conformal Killing vector. 
     It is easy to check that in $4, 5$ and $6$ dimensions
       the number of conditions coming from the vanishing of the gravitino and dilation variations, 
     i.e.\ the $d$-dimensional analogue of $\eqref{eq:L:BoundSusy}$, 
     is exactly the same as the number of conditions  corresponding to the existence of a conformal Killing vector
     plus the number of components of the bosonic auxiliary fields in the Weyl multiplet.
     
     Lat us discuss for example the 5d case. The generalised CKS equation of the gravitino imposes $4 \times 8$ conditions\footnote{
     Note that in $5d$, we have symplectic Majorana spinors.
     } and the dilatino brings another set of $8$, making a total of $40$ constraints coming from supersymmetry.
     This is to be confronted with the auxiliary bosonic background fields, which have a total of $26$ components \cite{LectParis}. 
     The $SU(2)$ gauge field has $15$ components,
     the scalar  $1$ and the tensor $T_{\mu \nu}$ additional $10$.
     The remaining $14$ degrees of freedom can be stored in a traceless symmetric degree $2$ tensor,
     corresponding to the      CKV condition , $\nabla_{(\mu} z_{\nu)} = \lambda g_{\mu \nu}$. 
The analysis is analogous in 6d. 
In some sense, the CKV condition takes the role of the degrees of freedom in the graviton, the tracelessness
being related to the Weyl invariance of the equation. The previous counting can be then reformulated as the equality of the fermionic and bosonic
off-shell degrees of freedom in the conformal gravity multiplet. The off-shell closure of the algebra is actually true only modulo gauge transformations 
and the previous argument should be taken as an analogy, although it probably can be made more precise.

The previous argument  suggests that  extended supersymmetry can be preserved on spaces with a   conformal Killing vector also in 5 and 6 dimensions.

       \section{The Euclidean Case}\label{sec:euclid}
       Supersymmetric theories on curved Euclidean manifolds have attracted much interest in the last years.
       This is mainly due to the possibility of calculating the partition function on some of these spaces,
       using supersymmetric localization techniques. 

       A consistent  supersymmetric definition of an  ${\cal N}=2$  theory  on any four-manifold 
       has been first introduced in \cite{Witten:1988ze}. There, supersymmetry is preserved by twisting the $SU(2)$ $R$-symmetry
       with one of the $SU(2)$'s of the Lorentz group $SO(4)$. The background gauge field precisely cancels the spin connection for a spinor 
       of definite chirality which becomes covariantly constant. In such a theory the energy-momentum tensor is 
       $Q$-exact and the quantum field theory is topological, i.e.\ the partition function and correlators are independent of the metric.
       In fact, the correlation functions compute the Donaldson polynomials of the four-manifold, which are topological invariants.
       In \cite{Pestun:2007rz} a different way of putting an $N=2$ theory on the four-sphere was introduced, essentially by conformal mapping from flat space,
       and the partition function has been computed through localization. In this case, the partition function
       was used to find a matrix model description for certain Wilson loops in the ${\cal N}=2$  theory. The spinor preserving supersymmetry
       is now a certain combination of (uncharged) CKS of opposite chirality on the sphere.  The construction has been generalised  in \cite{Hama:2012bg} where the same theories were put on a squashed four-sphere, preserving some supersymmetry.  The analysis allowed for a more detailed comparison with the $2$-dimensional duals via the AGT correspondence \cite{Alday:2009aq}\footnote{
       See also \cite{Hama:2013ama} where further details in this context have been studied.}. In this more general
       example background gauge and tensor fields take non-trivial values.
       
       In this section we want to discuss the geometrical constraints imposed by supersymmetry  and  determine
       general expressions for the background fields that we need to turn on.

       \subsection{Wick rotation of the conformal supergravity}
       In order to obtain the Euclidean supersymmetry conditions we have to Wick rotate $\eqref{eq:L:BoundSusy}$.
       Our strategy is to double the equations  and then impose a symplectic Majorana-Weyl condition on the spinors
\begin{align}
  &\left(\epsilon_+^i\right)^c = i \epsilon_{ij} \epsilon_+^j && \left(\epsilon_- {}_i\right)^c = -i \epsilon^{ij} \epsilon_-{}_j &
\end{align}
where $\left( \epsilon \right)^c \equiv B^{-1} \epsilon^*$. $\pm$ still denotes chirality. See the appendix \ref{sec:E:convent} for more details.
       We get two real equations for the gravitino
       \begin{equation} \label{eq:E:SUSY:gravitino} 
	   \begin{aligned} 
	      \nabla^A_m \epsilon^i_+ + \frac{i}{4} T^+_{m n} \gamma^n \epsilon_-^i - \frac14 \gamma_m D^A \epsilon_+^i &= 0 \\
	      \nabla^A_m \epsilon^i_- + \frac{i}{4} T^-_{m n} \gamma^n \epsilon_+^i - \frac14 \gamma_m D^A \epsilon_-^i &= 0 \, .
	   \end{aligned}
       \end{equation}
       Note that we have redefined $A_{m a}$ and $T^{\pm}_{mn}$ which are now both real.
       The vanishing of the dilatino  gives the two conditions
       \begin{equation}\label{eq:E:SUSY:dilatino} 
	   \begin{aligned} 
	      i \nabla^A_m T^+{}^m{}_n \gamma^n \epsilon_-^i +  D^A D^A \epsilon_+^i 
	     -  2 \nabla_m A_{n 4} \gamma^{m n} \epsilon_+^i + 2d \, \epsilon_+^i &=0 \\
		i \nabla^A_m T^-{}^m{}_n \gamma^n \epsilon_+^i + D^A D^A \epsilon_-^i 
	     +  2 \nabla_m A_{n 4} \gamma^{m n} \epsilon_-^i +2 d \, \epsilon_-^i &=0 \, .
	   \end{aligned}
       \end{equation}
       We have 
       \begin{align*}
	 \nabla^A_m \epsilon_+^i = \nabla_m \epsilon_+^i + \frac{i}{2} A_{ma} \bar \sigma^a{}^i{}_j  \epsilon_+^j &&
	 \nabla^A_m T^{+}_{mn} = (\nabla_m + A_{m4}) T^+_{mn} \\
	 \nabla^A_m \epsilon_-^i = \nabla_m \epsilon_-^i + \frac{i}{2} A_{ma} \sigma^a{}^i{}_j  \epsilon_-^j &&
	 \nabla^A_m T^{-}_{mn} = (\nabla_m - A_{m4}) T^-_{mn} 
       \end{align*}
       where we have defined 
       \begin{align} \label{eq:E:sigmas} 
	 \sigma^a{}^i{}_j = (\vec{\sigma}, i ){}^i{}_j  && 
	 \bar \sigma^a{}^i{}_j = (\vec{\sigma}, -i ){}^i{}_j  \, .
       \end{align} 

       After Wick rotation, $A_{m4}$ becomes an $SO(1,1)$ gauge field, the total $R$-symmetry being  $SO(1,1) \times SU(2)$.
       This is consistent with the R-symmetry group coming from compactifying $\mathbb{R}^{1,9}$ on $\mathbb{R}^6$ to four Euclidean dimensions, 
       see \emph{e.g.\ }\cite{Pestun:2007rz} and with the known classification of Euclidean superconformal algebras. The non-compactness of $A_{m4}$ is
       also necessary to make equations \eqref{eq:E:SUSY:gravitino} and \eqref{eq:E:SUSY:dilatino} consistent with the symplectic Majorana condition.

     Similarly, we can Wick rotate the Lagrangian and the supersymmetry transformations of arbitrary vector- and hypermultiplets 
     coupled to the Weyl multiplet above. 
     Here we note the form of the Lagrangian for a vector multiplet 
     \begin{multline} \label{eq:E:Lagrangian} 
       \mathcal{L}^{\text{Euclid}} = 
       d \, \phi \bar \phi + \nabla^A_m \phi \nabla^A{}^m \bar \phi + \frac{1}{8} Y^i{}_j Y^k{}_i - g \left[ \bar \phi , \phi \right]^2
	 +\frac{1}{8} F_{\mu \nu} F^{\mu \nu}  \\
	 + \frac{1}{4} \big( \bar \psi_-^i D^A \psi_+{}_i + \bar \psi_-{}_i D^A \psi_-^i\big) 
	 + \frac{i}{2}g \big( \bar \psi_-^i \left[ \phi, \psi_-{}_i \right]  - \bar \psi_+{}_i \left[ \bar \phi, \psi_+^i \right] \big) \\
	 -\frac14 \big( \phi F_{m n} T^+{}^{m n} + \bar \phi F_{m n} T^-{}^{m n} \big)
	 -\frac{1}{16} \big( \phi^2 T^+_{m n} T^+{}^{mn} + \bar \phi^2 T^-_{m n} T^-{}^{mn}  \big)
     \end{multline}
     where the Euclidean $SU(2)$ triplet satisfies $\left( Y^i{}_j \right)^* = - Y_i{}^j \equiv -\epsilon^{jk} Y_{ik}$.
     We will comment on the supersymmetry transformations below.

                     In the following we want to discuss on which manifolds we can solve the equations 
       $\eqref{eq:E:SUSY:gravitino}$ and $\eqref{eq:E:SUSY:dilatino}$. There is a degenerate situation where the spinor of one 
       chirality is set to zero. In this case the equations collapse to a non-abelian version of the CKS equation for the remaining spinor, which we can always
       solve by a twist,  as discussed later\footnote{We can not exclude that  more general solutions may exist.}. 
       From now on   we will focus on the general case where supersymmetry is preserved using spinors of both chiralities. 
       In this case  the result is very similar to the Lorentzian one:  the only geometrical constraint imposed by supersymmetry is the existence of a 
       CKV. We will solve explicitly the condition of supersymmetry and determine the auxiliary  fields for any manifold with a CKV. 

       \subsection{The geometry of spinors}
       It turns out that --albeit the technical details are quite different -- many of the equations that we have seen in the Lorentzian 
       describing the geometry defined by the spinors have a very close analogue in the Euclidean.
       In fact, two chiral Majorana-Weyl spinors define two scalars, two sets of real (anti-)selfdual two-forms 
       $\eta^x_{mn}$ and $\bar \eta^x_{mn}$ and again a tetrad $e^a_n$.
       To see this, note that we can construct the following spinor bilinears
       \begin{equation} \label{eq:E:Bilinears} 
	 \begin{aligned}
	   e^A &= \frac{1}{2} \epsilon_+^\dagger{}_i \epsilon_+^i \\
	   e^B &= \frac{1}{2} \epsilon_-^\dagger{}^i \epsilon_-{}_i   \\
	   e^a_m & = -\frac{e^{-b}}{2} \epsilon_-^{\dagger}{}_j \gamma_m \epsilon_+^i \bar \sigma^a{}^j{}_i \\
	   \bar \eta^x_{mn} &= -i\frac{e^{-A}}{2} \epsilon_{+}^\dagger{}_j \gamma_{mn} \epsilon_+^i \sigma^x{}^j{}_i \\
	   \eta^x_{mn} &= i\frac{e^{-B}}{2} \epsilon_{-}^\dagger{}_j \gamma_{mn} \epsilon_-^i \sigma^x{}^j{}_i 
	 \end{aligned} 
       \end{equation} 
       where $b=(A+B)/2$.
       As opposed to the Lorentzian, in the Euclidean all the forms are real. 
       The $\eta^x$ and $\bar \eta^x$ are (anti-)selfdual, respectively
       \begin{align*}
	 \epsilon_{mn}{}^{pq} \bar \eta^x_{pq} = -2 \bar \eta^x_{mn} &&
	 \epsilon_{mn}{}^{pq} \eta^x_{pq} = 2 \eta^x_{mn} \, .
       \end{align*}
     Applying the Fierz identities, one can show that, similarly to the Lorentzian case, the $e^a_n$'s form a tetrad,
     \begin{equation*}
       e^a \cdot e^b = \delta^{ab} \, .
     \end{equation*}
     In this frame, the two-forms have components
     \begin{align}
	   \bar \eta^x_{a b} = \delta_a^4 \delta_b^x -\delta_a^x \delta_b^4 + \epsilon^x{}_{a b} &&
	        \eta^x_{a b} =-\delta_a^4 \delta_b^x +\delta_a^x \delta_b^4 + \epsilon^x{}_{a b} \, 
     \end{align}
     and we give more details in the appendix.
     For completeness, let us mention how the forms defined by the spinors can be stored elegantly into the bispinors
       \begin{equation}
	 \label{eq:E:bispinors}
       \begin{aligned}
	 \epsilon_+^i \epsilon_-^\dagger{}_j &= -\frac{e^b}{4} \sum_a \left( e^a + \ast e^a \right) \sigma^{a}{}^i{}_j  \\
	 \epsilon_+^i  \epsilon_+^\dagger{}_j  &= \frac{i e^A}{4} \eta_a \sigma^a{}^i{}_j \\
	 \epsilon_-^i  \epsilon_-^\dagger{}_j  &= \frac{i e^B}{4} \bar \eta_a \sigma^a{}^i{}_j
       \end{aligned}
     \end{equation}
     where we have defined $\eta^4=-i(\gamma+1)$ and $\bar \eta^4 = i (\gamma-1)$.
     In this language, $e^a=e^a_m \gamma^m$, $\bar \eta_x = \frac12 \bar \eta_a{}_{mn} \gamma^{mn}$ and 
     $ \eta_x = \frac12 \eta_a{}_{mn} \gamma^{mn}$. 
     The hodge dual of a one-form is $\ast e^a = \gamma e^a$.

     As in the Lorentzian case, we will mainly work in the frame defined by the $e^a$'s.
     Again, the action of the flat gamma matrices can then be translated into multiplication with Pauli matrices, 
       \begin{equation} \label{eq:E:gammaxeps} 
	 \begin{aligned}
	 \gamma^a \epsilon_+^i &= - e^{\Delta}  \sigma^a{}^i{}_j \epsilon_-^j &\quad \quad \quad \quad \quad &
	 \gamma^{ab} \epsilon_+^i = -i \bar \eta_x^{ab} \sigma^x{}^i{}_j \epsilon_+^j \\
	 \gamma^a \epsilon_-^i &= - e^{-\Delta} \bar \sigma^a{}^i{}_j \epsilon_+^j &&
	 \gamma^{ab} \epsilon_-^i = -i  \eta_x^{ab} \sigma^x{}^i{}_j \epsilon_-^j 
       \end{aligned}
       \end{equation}
     where $\Delta = (A-B)/2$.
     This can be used to show that in this frame the spinors are constant (up to an overall norm factor).

       Before we discuss the supersymmetry conditions in the next section, note that a convenient base for spinor doublets 
       of positive and negative chirality is given by, respectively,
       \begin{align} \label{eq:E:SpinorBase} 
	 \bar \sigma^a{}^i{}_j \epsilon_+^j  && \sigma^a{}^i{}_j \epsilon_-^j \, .
       \end{align} 
       Since the spinors are Majorana-Weyl, each of the two bases contains $4$ real components.

      We define the intrinsic torsions
       \begin{align} \label{eq:E:DefP} 
	 \nabla_m \epsilon_+^i = -\frac{i}{2} \bar P_{m a} \bar \sigma^a{}^i{}_j \epsilon_-^j &&
	 \nabla_m \epsilon_-^i = -\frac{i}{2} P_{m a} \sigma^a{}^i{}_j \epsilon_+^i 
       \end{align} 
       where $P_{m a}$ and $\bar P_{m a}$ are independent real objects. 
       In the frame defined by the spinors, they are, as in the Lorentzian case, composed of the spin connection and the norms of the spinors
     \begin{equation} \label{eq:E:Torsions} 
       \begin{aligned} 
	 \bar P_{m4} &= -\partial_m A &\hspace{3cm}&
	 \bar P_{mx} = \frac12 \omega_m{}^{ab} \bar \eta_x{}_{ab} \\
	  P_{m4} &= \partial_m B &&
	  P_{mx} = \frac12 \omega_m{}^{ab} \eta_x{}_{ab} 
       \end{aligned} 
     \end{equation}
     Here $\eta_x{}_{ab}$ and $\bar \eta_x{}_{ab}$ project on the (anti-)selfdual part of the spin connection, respectively.

    Our choice of frame degenerates at the points where the spinors vanish.  A more detailed analysis will be necessary at the  vanishing locus
    of the spinors to ensure that the results we will obtain are globally defined. 

     We are now ready to classify the solutions to the equations $\eqref{eq:E:SUSY:gravitino}$ and $\eqref{eq:E:SUSY:dilatino}$.

     \subsection{Solving the supersymmetry conditions} \label{sec:E:ClassGeo}
     The analysis will be very similar to the Lorentzian one in section \ref{sec:L:ClassGeo}.
     Again we define the symmetric traceless part of the torsions
       \begin{align*}
	 p_{a b}  \equiv P_{(a b)} - \frac14 \delta_{a b} P_c^c &&
        \bar p_{a b}  \equiv \bar P_{(a b)} - \frac14 \delta_{a b} \bar P_c^c
       \end{align*}
     and it is an easy exercise to check that the gravitino equation $\eqref{eq:E:SUSY:gravitino}$ is solved by
       \begin{subequations}
	 \label{eq:E:solutionGravitino}
         \begin{align}
	   \bar p_{a b} &= p_{ab}   \label{eq:E:CKSp}\\
	    p_{a b} &= A_{(a b)} - \frac{\delta_{a b}}{4} A_c^c  \label{eq:E:ASTrless}\\
	   T^+_{a b} &= -2 e^{\Delta} \left( \bar P^+_{a b} -  A^+_{a b} \right) \label{eq:E:Tplus} \\
	   T^-_{a b} &= -2 e^{-\Delta} \left(  P^-_{a b} -  A^-_{a b} \right) \label{eq:E:Tminus} \, .
         \end{align}
       \end{subequations}
     The first equation tells us that $z=e^b e^4$ is a conformal Killing vector, 
     while the other three equations fix some parts of the background fields.
     More explicitly, the conformal Killing condition $\nabla_{(m} z_{n)} = \lambda g_{mn}$
     implies
     \begin{align}
       &\lambda = e^{b} \partial_4 b 
       &
       & \omega_{(x}{}^{4}{}_{y)} = -\delta_{xy} \partial_4 b &
       & \omega_4{}^4{}_x = \partial_x b
       \label{eq:E:CKVomega}
     \end{align}
     where we have denoted $\partial_4 = e^m_4 \partial_m$ etc.
     It is an easy task to see that these conditions are equivalent to $\eqref{eq:E:CKSp}$, 
     using the explicit formulae given in $\eqref{eq:E:Torsions}$.
     The other three equations in $\eqref{eq:E:solutionGravitino}$ determine the ``symmetric traceless" part of the gauge field
     and the value of the tensor field, respectively.
     As in the Lorentzian case, it still remains to determine the ``antisymmetric" part of the gauge field, its ``trace" and the scalar $d$.
     In close analogy, this is done by the dilatino equation, which does not impose any new restrictions on the geometry.
     
     Similarly to the Lorentzian case, we introduce a re-definition  of the gauge field 
      \begin{equation} \label{eq:E:solA}
      \begin{aligned}
       A_{x4} &\equiv -b_x - \partial_x (b+\Delta) \\
       A_{4x} &\equiv b_x + \frac12 \epsilon_{xyz} \omega_4{}^{yz} + \partial_x b \\
       A_{xy} &\equiv \epsilon_{xy}{}^z a_z + \frac14 \delta_{xy} \alpha + \frac12 \epsilon_{uvy} \omega_{x}{}^{uv} + \omega_{[x}{}^4{}_{y]} \\
       A_{44} &\equiv \frac14 \alpha -\partial_4 \Delta\, .
      \end{aligned}
     \end{equation}
     Note that $a_x$ and $b_x$ correspond to the anti-symmetric part of the twisted torsion $\bar P$
     \begin{align} \label{eq:E:Pshifted}
       \bar P_{[4x]} - A_{[4x]} = -b_x &&
       \bar P_{[xy]} - A_{[xy]} = -\epsilon_{xy}{}^z a_z
     \end{align}
     and, in terms of this re-definition, the tensor field in $\eqref{eq:E:solutionGravitino}$ can be written as
     \begin{equation} \label{eq:E:TShifted} 
       \begin{aligned}
	 e^{-\Delta} T^+_{4x} &= b_x +a_x \\
	 e^{\Delta} T^-_{4x} &= b_x -a_x +2\partial_x b -\epsilon_x{}^{yz} \omega_y{}^4{}_z \, .
       \end{aligned}
     \end{equation}
     Note that $2 \partial_x b -\epsilon_x{}^{yz} \omega_y{}^4{}_z = -4 \left( \nabla_{[4} z_{x]} \right)^- = \eta_x{}^{ab} \nabla_a z_b $. 
     The remaining components of $T^+$ and $T^-$ are fixed according to self-duality.
     
     The analysis of the dilatino equation $\eqref{eq:E:SUSY:dilatino}$ is similar to the Lorentzian case.
     It can be re-written in a way that it fixes the seven missing pieces of the gauge field  
     \begin{equation} \label{eq:E:DilatinoResult} 
       \begin{aligned}
	 \partial_4 a_x + (\omega_x{}^y{}_4 - \omega_4{}^y{}_x) a_y &= e^{-b}\partial_x (e^b \partial_4 b)  \\
	 \partial_4 b_x + (\omega_x{}^y{}_4- \omega_4{}^y{}_x ) b_y &=-e^{-b}\partial_x (e^b \partial_4 b)  \\
	 (\partial_4 +\partial_4 b) \alpha &= 0
       \end{aligned}
     \end{equation}
     and the scalar field $d$
     \begin{equation} \label{eq:E:DilatinoResult:4}
     \begin{aligned}
	 d&= -\partial_4 \partial_4 b - 2\left( \partial_4 b \right)^2 + \frac{1}{16} \alpha^2 &  \\
	 &- (\partial_x + \omega_z{}^z{}_x -\frac12 \epsilon_{x}{}^{yz} \omega_y{}^4{}_z -\frac12 a_x )  a^x
	 + \frac12 (\epsilon_x{}^{yz} \omega_y{}^4{}_z +2 \partial_x b + b_x) b^x  
       \end{aligned}
     \end{equation}
     In these equations, 
     expressions like $\partial_x \partial_y b$ are to be understood as $e_x^m \partial_m (e_y^n \partial_n b)$ etc.

       We can again pick a set of local coordinates that automatically solve $\eqref{eq:E:DilatinoResult}$ and 
       leave us with an arbitrariness in the ``anti-symmetric" and the ``trace"-part of the gauge field.
       After a Weyl rescaling and a choice of coordinates such that $z=\partial/\partial \xi$, the metric can locally be written as 
       \begin{align} \label{eq:E:MetricKilling} 
	 ds^2 = e^{2b} (d\xi + \mathcal{F})^2 + \mathcal{H}_{ij} dx^i dx^j
       \end{align} 
       where $b$, $\mathcal{F}$ and $\mathcal{H}$ do not depend on $\xi$. 
       $\mathcal{F}$ is  a one-form on the ``spatial" part transverse to $z$.
       As a one-form, we have 
       \begin{equation} \label{eq:E:zoneform} 
	 z= e^{2b}(d\xi+\mathcal{F}) \, .
       \end{equation} 
       In such coordinates we can have additional symmetries of the spin connection. 
       By choosing the three-dimensional frame $e^x$ such that $e^4 \cdot de^x = 0$, one gets
       \begin{equation} \label{eq:E:symmofspinconnection} 
	 \omega_4{}^y{}_x = \omega_x{}^y{}_4 
       \end{equation} 
       and $\eqref{eq:E:DilatinoResult}$ boils down to
       \begin{align} \label{eq:E:DilatinoResultCoord} 
	 \partial_\xi \alpha = \partial_\xi a_x = \partial_\xi b_x &= 0  \, .
       \end{align} 
       We see that also in the Euclidean case we have the freedom to choose arbitrary values for $a_x, b_x$ and $\alpha$
       as long as they do not depend on the isometry $\xi$.
       
       For example  we can always use the freedom in the gauge field to locally impose $T^+_{mn} = T^-_{mn} = 0$.
From $\eqref{eq:E:TShifted}$ we see that this condition is satisfied for
\[
a_x=-b_x= \partial_x b - \frac{1}{2} \epsilon_{x}{}^{yz} \omega_y{}^{4}{}_{z} \, 
\]
and it is a quick check that these $a_x$ and $b_x$ solve $\eqref{eq:E:DilatinoResult}$, where one has to use the Bianchi identities.
In the coordinates introduced in $\eqref{eq:E:MetricKilling}$, we find for the gauge and scalar fields
\begin{equation} \label{eq:E:AwithTzero} 
  \begin{aligned}
    A_{44} &=  \frac{\alpha}{4} \\
    A_{x4} &= -\big( \partial_x \Delta + e^b \left( \tilde \ast \, d\mathcal{F} \right)_x \big) \\
    A_{4x} &=  2 e^b \left( \tilde \ast \, d\mathcal{F} \right)_x\\
    A_{xy} &= \epsilon_{xy}{}^z \partial_z b + \frac{1}{2} \epsilon_{uvy}  \omega_x{}^{uv} + \frac{\delta_{xy}}{4} \alpha \, \\
  \end{aligned}
\end{equation} 
where $\tilde \ast$ denotes the three dimensional Hodge dual. 
The value for the scalar $d$ follows from $\eqref{eq:E:DilatinoResult:4}$.

       To summarize, in complete analogy to the study of the Lorentzian case, the only constraint imposed by  supersymmetry is the existence of 
       a conformal Killing vector. 
       Given such a vector $z \cdot z = e^{2b}$, we can preserve some supersymmetry
       by turning on background values for the $SO(1,1) \times SU(2)$ gauge field as determined in $\eqref{eq:E:solA}$
       and for the tensor field as in $\eqref{eq:E:TShifted}$,
       where $a_x, b_x, \alpha$ and $\Delta$ are free parameters of the solution subjected to $\eqref{eq:E:DilatinoResult}$. 
       One also has to turn on a background scalar field as in $\eqref{eq:E:DilatinoResult:4}$. 

The previous expressions may become singular at the points where the spinors vanish, in particular where the conformal Killing vector degenerates.
A more careful analysis is required near the zeros of the spinors in order to ensure that the solution is  regular. The large arbitrariness
in the choice of auxiliary fields should usually allow to find globally defined solutions.

       \subsection{Examples}
In this section we present some examples for the formalism we have introduced above. 


       \subsubsection{Round and squashed spheres}
       Round and squashed spheres have been a main focus in the study of 
       supersymmetry on curved spaces and exact results in quantum field theory.
       In fact, the work of Pestun \cite{Pestun:2007rz}, who considered $\mathcal{N}=2$ theories on the round $S^4$,
       has in a sense triggered the recent activity in this field.
       This work has been generalised in \cite{Hama:2012bg}, where $\mathcal{N}=2$ theories on a squashed sphere, or ellipsoid,  were considered.
       The squashing preserves a $U(1) \times U(1)$ isometry and the manifold  can be described by the equation
       \begin{align*}
	 \frac{x_1^2+x_2^2}{\ell} + \frac{x_3^2 + x_4^2}{\tilde \ell} + \frac{x_5^2}{r^2} = 1 \,  
       \end{align*}
       where $\ell$ and $\bar \ell$ are the squashing parameters and $r$ is the radius of the sphere.
       A metric for this space is given by
       \begin{equation} \label{eq:E:EllipsoidMetric} 
	 (g^2+h^2) d\rho^2 + 2 f h \sin\rho \, d\theta \, d\rho  + 
	 \sin^2 \rho \,  (f^2 d \theta^2 + \ell^2 \cos^2\theta \, d\phi^2 + \tilde \ell^2 \sin^2 \theta \, d\chi^2 ) 
       \end{equation} 
       where the functions $f,g$ and $h$ are defined in \cite{Hama:2012bg}
       \begin{equation} \label{eq:E:Ellipsoidfgh} 
	 \begin{aligned}
	   f &= \sqrt{\ell^2 \sin^2 \theta  + \tilde \ell^2 \cos^2 \theta } \\
	   h &=  (\tilde \ell ^2 - \ell^2 )f^{-1} \cos \theta \sin \theta \cos \rho\\
	   g &= \sqrt{\ell^2 \tilde \ell^2 f^{-2} \cos^2 \rho +r^2 \sin^2 \rho}
	 \end{aligned}
       \end{equation}
       On the ellipsoid there is a Killing vector (here identified with the dual one-form) 
       \[
       z = \frac{1}{\ell} \partial_{\phi} + \frac{1}{\tilde \ell} \partial_\chi \, 
       \hat =\,  \sin^2 \rho \, (\ell \cos^2 \theta \, d \phi + \tilde \ell \sin^2 \theta\, d\chi)\,.
       \]
      with $z \cdot z = e^{2b} = \sin^2 \rho$.
       In order to apply the formulae discussed in this paper, we choose a frame with $ e^4 = e^{-b} z$,
	 \begin{equation} \label{eq:E:EllipsoidFrame} 
	   \begin{aligned}
	     e^1 &= f \sin \rho \, d\theta + h d \rho \quad \quad \quad
	     &e^3 &= \sin \rho \, \cos \theta \sin \theta  (\ell d\phi - \tilde \ell d \chi) \\
	     e^2 &= g d\rho 
	     &e^4 &= \sin^{-1} \rho \; z
	   \end{aligned}
	 \end{equation}
       which is related to the vierbein in \cite{Hama:2012bg} by an $SO(4)$ rotation. 
       Let us discuss the round and the squashed sphere separately.

       The round sphere result of Pestun, with all background fields but the scalar vanishing,
       is recovered in the limit $\ell = \tilde \ell = r$.
       In particular, it should be reproduced by our special solution with vanishing tensor field described  around $\eqref{eq:E:AwithTzero}$.
       Plugging the explicit frame $\eqref{eq:E:EllipsoidFrame}$ into $\eqref{eq:E:AwithTzero}$,
       we find that the vector field is pure gauge, $A^i{}_j = - (d\phi+d\chi) \sigma^2{}^i{}_j$, when  $\Delta = \log \cot \frac{\rho}{2}$ 
       and $\alpha=0$.
       Note that $SU(2)$ gauge transformations correspond to local Lorentz transformations in the three ``spatial" coordinates,
       as follows from the definition of the vierbein in $\eqref{eq:E:bispinors}$.
       In fact, if we rotate the frame $\eqref{eq:E:EllipsoidFrame}$ by $(\phi+\chi)$ around $e^2$, 
       we find that the gauge field vanishes identically.
       The scalar field $\eqref{eq:E:DilatinoResult:4}$ is constant, $d=2/r^2=R/6$ \footnote{This value corresponds to 
       a vanishing scalar $\tilde d = 0$ in the original Weyl multiplet $\eqref{eq:Weylmplt}$.}.
       
       Now let us discuss the case of the ellipsoid. The solution found in  \cite{Hama:2012bg} has non-vanishing tensor fields $T^\pm$
       and an $SU(2)$  gauge field $A_{\mu x}$ whose explicit expressions can be found in equations (3.28) and (3.29) of that paper.
       Of course, the same result follows from our formalism.   After computing the spin connection from the vierbein $\eqref{eq:E:EllipsoidFrame}$,
       all background fields are fixed in terms of this geometrical data, plus the eight free parameters in our solutions.
       This is to be contrasted with the result  in \cite{Hama:2012bg}, which is a $3$-parameter family instead.
       The reason for the mismatch is that the authors of \cite{Hama:2012bg} work with explicit spinors and 
       have switched off the $SO(1,1)$ gauge field.
       We checked that choosing $\Delta$ appropriately and tuning the $SO(1,1)$ vector to zero 
       indeed leads to their $3$-parameter solution.
       More explicitly, setting $A_{m4} = 0$ eliminates four of the parameters,
	 \begin{align}\label{eq:E:EllipsoidSolution} 
	   \alpha = 0 &&
	   b_x = \delta_{x2} \frac{1}{g} \tan \frac{\rho}{2} 
	 \end{align}
         where we set the value of $e^\Delta = \cot \frac{\rho}{2}$.
	 Our scalar $\eqref{eq:E:DilatinoResult:4}$, the gauge field $\eqref{eq:E:solA}$ and the tensor $\eqref{eq:E:TShifted}$
	 are then identical\footnote{
	 Note that there is a reshuffling in the order of $SU(2)$ indices. 
	 Furthermore, the two gauge fields differ by a gauge transformation.
	 } to the ones in \cite{Hama:2012bg}, when we identify our parameters $a_x$ with their $c_1, c_2$ and $c_3$,
           \[
	   a_x= \delta_{x1} \left( 4 c_2 -\frac{h}{fg} \tan \frac{\rho}{2}  \right) 
	   + \delta_{x2} \left(  4c_1 - \frac{1}{f} \tan \frac{\rho}{2} \right)
	   - \delta_{x3} 4 c_3 \, .
	   \]
	   
	   The three parameters $c_i$ can be chosen as in  \cite{Hama:2012bg} in order to have a regular solution on $S^4$ which reduces
	   at the North and South pole of the squashed sphere to the $\Omega$-background  \cite{Nekrasov:2003rj} which plays an important role in 
	   reducing the computation of the partition function to the Nekrasov partition functions for instantons. The local form of this solution and the $\Omega$-background are discussed in the next section.
       
       \subsubsection{Twisting the theory}
       In this subsection we want to discuss the special case $\Delta = -b$.
       This choice is is actually related to a general solution discussed in \cite{Nekrasov:2003rj}.
       There the authors have noticed that it is possible to preserve extended supersymmetry
       on every manifold with a Killing vector $z$ by generalising
       the Witten twist for $\mathcal{N}=2$ theories \cite{Witten:1988ze}.

       Let us first discuss  the Witten twist in our formalism. The twist corresponds to the degenerate situation where  
       one uses only one  chiral spinor, say $\epsilon_+$,  while the spinor of the other chirality vanishes and  we cannot immediately
       use our previous formulae. 
       Since only the self-dual
       part of the spin-connection acts on $\epsilon_+$, corresponding to an $SU(2)$ subgroup of the tangent group $SO(4)$,
       the spinor can be made covariantly constant by cancelling the spin-connection with the 
       $SU(2)$-gauge field in the covariant derivative. More explicitly, a
       constant spinor $\partial_m \epsilon_+^i = 0 $ will be covariantly constant  if we turn on the  
       $SU(2)$-gauge field 
       \begin{equation} \label{eq:E:ATwistSolution}
	 A_{mx} = \frac12 \omega_m{}^{ab} \bar \eta_x{}_{ab} \, .
       \end{equation}
       The generalised CKS equation is then satisfied with vanishing tensor fields.
       One can check that the dilatino equation is also satisfied with  a value for the scalar field $\tilde d=-R/6$. 
       This computation was first done long time ago in \cite{Karlhede:1988ax}.        
       
       Let us now consider a more general case, where the gauge field is still as in $\eqref{eq:E:ATwistSolution}$ but
        the negative chirality spinor is different from zero.
       As suggested in \cite{Nekrasov:2003rj}, 
       the pair $(\epsilon_+^i, \epsilon_-^i = i \gamma^a z_a \epsilon_+^i)$ preserves supersymmetry on any manifold with Killing vector $z$.
       We can easily see this in the picture of coupling to conformal supergravity discussed in this paper. 
       As it is clear from equation $\eqref{eq:E:gammaxeps}$, the relation 
       $\epsilon_-^i = i \gamma^a z_a \epsilon_+^i$ is true for our spinors exactly when the norm of $\epsilon_+$ vanishes, or
       \[
       \Delta = -b \,.
       \]

       One can then  check
       that for the choice of background fields 
        \begin{align}\label{backs}
	  d = 0 && T^+_{ab} = 0 && T^-_{ab} = -4 e^b \left( \nabla_{[a} z_{b]} \right)^- \, 
       \end{align}
       both $\epsilon_+$ and $\epsilon_- = i z \epsilon_+$ 
       are generalised conformal Killing spinors fulfilling $\eqref{eq:E:SUSY:gravitino}$ and $\eqref{eq:E:SUSY:dilatino}$.
       There is obviously an analogous solution with a covariantly constant spinor $\epsilon_-^i$ and a self-dual tensor $T^+$.

       For completeness, let us derive it from our general expressions above.
       For the choice $ \Delta = -b$, our gauge field takes the value
       \begin{align*}
	 A_{m4} = b_m  && A_{mx} = \frac12 \omega_m{}^{ab} \bar \eta_x{}_{ab} + \delta_m^y \epsilon_{yx}{}^z a_z \, 
       \end{align*}
       where we have called $b_4 \equiv \alpha  +4\partial_4 b$.
	Recall that the covariant derivative of $\epsilon_+$ is
	\begin{equation}
	      \nabla_a^A \epsilon_+^i = \frac{i}{2} \bar P_{ab}^A \gamma^b \epsilon_-^i \, .
	\end{equation}
       If we set $b_4 = 0$ by an appropriate choice of $\alpha$, the torsion $\bar P$ becomes anti-symmetric, $\bar P_{mn} = \bar P_{[mn]}$,
       and is given by linear expressions in $a_x$ and $b_x$ as in $\eqref{eq:E:Pshifted}$.
       So we see that for vanishing $a_x$ and $b_x$ the spinor $\epsilon_+$ is covariantly constant 
       and the gauge field cancels the self-dual part of the spin-connection as in $\eqref{eq:E:ATwistSolution}$.
       Hence, the class of our solutions with $\Delta = -b$ and $b_m = a_x =0$ reduces to the generalised twist solution 
       described above. The value of the scalar $\eqref{eq:E:DilatinoResult:4}$ and the tensor field $\eqref{eq:E:TShifted}$ become as in \eqref{backs}.
      \subsubsection{The $\Omega$-background} 
      One notable example of the Nekrasov-Okounkov twist discussed in the previous section is the $\Omega$-background on flat $\mathbb{R}^4$. It uses the Killing vector
      \begin{equation}            
      z = \epsilon_1 \left( x_1\partial_{x_2} -x_2\partial_{x_1}\right )  + \epsilon_2 \left (x_3\partial_{x_4} -x_4\partial_{x_3}\right ) \nonumber
      \end{equation} 
     An ${\cal N}=2$ theory will still preserve some supersymmetry on flat space if the tensor field 
     \begin{equation}
     T^- =  -2 (\epsilon_1+\epsilon_2) (dx_1 dx_2 + dx_3 dx_4 )\nonumber
     \end{equation}
     is turned on.  The corresponding field theory has a prepotential  determined in terms of the   Nekrasov instanton partition function.
     Alternatively, in the analogue solution with a covariantly constant negative chirality spinor, supersymmerty requires turning on  
     \begin{equation}
     T^+ =  2 (\epsilon_1-\epsilon_2) (dx_1 dx_2 - dx_3 dx_4 )\nonumber \, .
     \end{equation}
     This example is discussed at length in \cite{Hama:2012bg} where the background fields on the squashed sphere  have been chosen
     in order to reduce to the $\Omega$-background near the poles, with tensors of opposite chirality at the North and South pole.

     \section{Conclusions}

     In this paper we  have determined the background fields that generate supersymmetric curvature couplings for $N=2$ theories on Lorentzian and Euclidean 
     four-manifolds with a conformal Killing vector. The results can have interesting applications to holography and localization.
     
     We haven't discussed in detail the holographic part of the  story. As it happens for ${\cal N}=1$ theories \cite{Cassani:2012ri}, all the supersymmetric asymptotically AdS
     solutions of  ${\cal N}=4$ gauged supergravity in five dimensions will reduce to supersymmetric boundary theories which fit in the classification
     we have discussed. It would be interesting to find explicit regular bulk solutions that are dual to ${\cal N}=2$ boundary theories on curved space.

     In the Euclidean case one interesting application of the field theory in curved space is supersymmetric localisation.
     Typically, the functional distinguishing the locus for the localised path integral is
     quadratic in the fermion variations \cite{Pestun:2007rz, Hama:2012bg}
     \begin{align*}
       QV = \overline{\delta \psi_+^i} \delta \psi_+^i + \overline{\delta \psi_-^i} \delta \psi_-^i \, .
     \end{align*}
     In our formalism, the Euclidean gaugino variation in the vector multiplet reads\footnote{We define the barred variations in a way to render $QV$ positive definite
     by complex conjugation with the unphysical reality condition $\phi^\ast = \bar \phi$.
     This amounts to deforming the saddle point of the path integral contour into the complex plane \cite{Pestun:2007rz}.}
     \begin{align*}
       \delta \psi_+^i &= i D \phi \epsilon_-^i 
       - \frac14 \big( 
       (F^+_{mn} + e^\Delta \bar \phi \bar P^A{}^+_{mn} + e^{-\Delta}  \phi P^A{}^+_{mn}) \gamma^{mn} + e^{-\Delta} \phi P^A{}_m^m 
       \big) \epsilon_+^i 
       +\frac12 Y^i{}_j \epsilon_+^j - \left[ \phi, \bar \phi \right] \epsilon_+^i \\
       \delta \psi_-^i &= - i D \bar \phi \epsilon_+^i 
       + \frac14 \big( 
       (F^-_{mn} + e^\Delta \bar \phi \bar P^A{}^-_{mn} + e^{-\Delta}  \phi P^A{}^-_{mn}) \gamma^{mn} - e^{\Delta} \bar \phi \bar P^A{}_m^m 
       \big) \epsilon_-^i 
       -\frac12 Y^i{}_j \epsilon_-^j - \left[ \phi, \bar \phi \right] \epsilon_-^i \, .
     \end{align*}
     We leave a detailed analysis of the solution to $QV=0$ and applications to localisation to future work.
     We expect that there will be always a locus with all fields but the scalar $\phi$ vanishing.
     If the supersymmetry spinor -- and hence the conformal Killing vector of the geometry --
     has some zeros, the path integral typically gets contributions also from
     small instantons localised at the points where the Killing vector degenerates.
     In \cite{Pestun:2007rz,Hama:2012bg} the authors analysed cases where at these points the geometry locally looks like the 
     $\Omega$-deformation of flat space and hence the instanton contribution is given by the Nekrasov partition function \cite{Nekrasov:2002qd}.
     It would be interesting to study more general field theories on curved spaces and even cases with a non-degenerate Killing vector
     to see if in that case the localised partition function is exempt from any non-perturbative contribution.
     
     Finally, it would be interesting to analyse superconformal theories in five or six dimensions
     where the minimal supersymmetry has a structure similar to that discussed in this paper.
     The analysis of supersymmetry  in curved space may shed some light on the properties of the still elusive superconformal fixed points in five or six dimensions.
     As mentioned in section \ref{sec:higher} it is plausible that the condition of supersymmetry  in $d$-dimensions, at least in the Lorentzian case, will just reduce to the existence of a CKV, all other constraints 
     determining the auxiliary background fields.

\section*{Acknowledgments}
We would like to thank  D.~Klemm for interesting discussions and in particular A.~Tomasiello for initial collaboration and many insightful comments. 
The authors are supported in part by INFN, by the MIUR-FIRB grant RBFR10QS5J ``String Theory and Fundamental Interactions'', and by the MIUR-PRIN contract 2009-KHZKRX.

\appendix

\section{Spinors \& Conventions} \label{sec:convent}
Throughout the paper Greek indices denote Lorentz signature. 
The indices $\mu, \nu,\dots$ are curved and $\alpha, \beta, \dots = 0,1,2,3$ are flat.
Similarly, we use Roman indices for Euclidean spaces. More explicitly, $m, n,\dots $ are curved while $a,b,\dots=1,2,3,4$ are flat.
We denote with $x,y,\dots$ flat ``spatial" indices running from $1$ to $3$, both in Euclidean and in Lorentzian signature. 
\subsection{Lorentzian Signature} \label{sec:L:convent}
We work with a mostly plus signature, $(\eta_{\alpha \beta}) = (-,+,+,+)$.
The conventions for our gamma matrices are
\begin{align}
  &\gamma_\mu^* = \gamma_\mu  && \gamma_\mu^\dagger = \gamma^0 \gamma_\mu \gamma^0 &
  & \gamma = i \gamma^0 \gamma^1 \gamma^2 \gamma^3 &&  \gamma^* = \gamma^T = -\gamma \;.
  \label{eq:L:gammas}
\end{align}
Following the conventions in \cite{LectParis}, we have supersymmetry parameters that are chiral spinor doublets of $SU(2)$
\begin{align}
  &\epsilon_+^i && \epsilon_- {}_i = \left( \epsilon_+^i \right)^* &
  & \eta_+{}_i && \eta_-^i = \left( \eta_+ {}_i\right)^*  \;
  \label{eq:L:susyparam}
\end{align}
and we define
\begin{align}
  &\bar \epsilon_+{}_i = \left(\epsilon^i_+\right)^\dagger \gamma^0 && \bar \epsilon_-^i = \left(\epsilon_-{}_i\right)^\dagger \gamma^0 \; .
\end{align}
We can use the $SU(2)$ invariant  tensor to raise and lower indices 
\begin{align} \label{eq:L:susyparamlower}
  & \epsilon_{+i} = \epsilon_{ij} \epsilon_+^j && \epsilon_-^i = \epsilon^{ij} \epsilon_{-j} &
  & \eta_+^i = \epsilon^{ij} \eta_{+j} && \eta_{-i} = \epsilon_{ij} \eta_-^j \, ,
\end{align}
where $\epsilon_{12} = \epsilon^{12} = 1$. 
Note that this is different to the conventions in \cite{LectParis}.
As opposed to there, for us the $SU(2)$ index position does not denote chirality but the $SU(2)$ representation:
we allow for raising and lowering with $\epsilon_{ij}$ and put explicit labels $+/-$ to indicate chirality.
The reader who wants to compare with \cite{LectParis} should contract with $\epsilon_{ij}$ whenever the
index structure of a spinor here is different from its analogous spinor there.

In our conventions, the (anti-)selfduality conditions read
\begin{align}
  \epsilon_{\mu \nu}{}^{\rho \tau} \Omega^{\pm}_{\rho \tau} = \pm 2 i \Omega^{\pm}_{\mu \nu}  \; .
  \label{eq:L:selfduality}
\end{align}
Note that $\left( \Omega^+ \right)^* = \Omega^-$.

       Let us also summarize some details on the  symbols $\eta^x$.
       They enter our equation through the identity
       \[
	 \sigma^m \sigma^n = -\eta_{xm}{}^n \sigma^x + \delta_m^{n} \mathbb{1}  \, 
       \]
       and they obey the following orthogonality and self-duality relations
       \begin{equation} \label{eq:L:eta}
       \begin{aligned}
	  \eta_x{}_{\mu \nu} \eta_x^{\rho \tau} &= i \epsilon_{\mu \nu}{}^{\rho \tau} -2 \delta_{[\mu}{}^\rho \delta_{\nu]}{}^\tau 
	  & \hspace{1.8cm} \eta_x{}_{\mu \nu} \eta_y^{\mu\nu} &= -4 \delta_{xy}  \\
	  \epsilon_{xyz} \eta_y{}_{\mu \nu} \eta_z^{\rho \tau} &= -4 i \delta_{[\mu}{}^{[\rho} \eta_x{}_{\nu]}{}^{\tau]} 
	  &\epsilon_{\mu \nu}{}^{\rho \tau} \eta_x{}_{\rho \tau} &= 2i \eta_x{}_{\mu \nu}
       \end{aligned}
     \end{equation}
       where $\epsilon_{xyz} = \epsilon_{0xyz}$ and $\epsilon_{0123}=1$.

\subsection{Euclidean Signature}  \label{sec:E:convent}
We work with gamma matrices
\begin{equation} \label{eq:E:gammas}
\begin{aligned}
  \gamma_m^* &= \gamma_m^T = B \gamma_m B^{-1}  &\hspace{1.5cm} & \text{with} \; \; \; B^* = B^T = -B^{-1} = -B & \\
   \gamma &= \gamma^1 \gamma^2 \gamma^3 \gamma^4 &&  \quad \quad \quad  \gamma^* = \gamma^T = \gamma \; .
\end{aligned}
\end{equation}
The supersymmetry parameters are symplectic Majorana-Weyl
\begin{align}
  &\left(\epsilon_+^i\right)^c = i \epsilon_{ij} \epsilon_+^j && \left(\epsilon_- {}_i\right)^c = -i \epsilon^{ij} \epsilon_-{}_j &
  & \left(\eta_+{}_i\right)^c = -i\epsilon^{ij} \eta_+{}_j  && \left(\eta_-^i\right)^c = - \epsilon_{ij} \eta_-^j 
  ,\label{eq:E:susyparam}
\end{align}
where $\left( \epsilon \right)^c \equiv B^{-1} \epsilon^*$ and the position of the $SU(2)$ index distinguishes the two representations.
We also define 
\begin{align}
  \epsilon_+^\dagger{}_i = \left( \epsilon_+^i \right)^\dagger &&  \epsilon_-^{\dagger i} = \left( \epsilon_-{}_i \right)^\dagger
\end{align} 
Again, $\epsilon_{ij}$ is used to raise and lower indices 
\begin{align}
  & \epsilon_{+i} = \epsilon_{ij} \epsilon_+^j && \epsilon_-^i = \epsilon^{ij} \epsilon_{-j} &
  & \eta_+^i = \epsilon^{ij} \eta_{+j} && \eta_{-i} = \epsilon_{ij} \eta_-^j \, ,
  \label{eq:E:susyparamlower}
\end{align}
where $\epsilon_{12} = \epsilon^{12} = 1$.

We use conventions with (anti-)selfduality conditions as
\begin{align}
  \epsilon_{m n}{}^{p q} \Omega^{\pm}_{p q} = \mp 2  \Omega^{\pm}_{m n}  \; .
  \label{eq:E:selfduality}
\end{align}

       The t'Hooft symbols $\eta$ in Euclidean signature appear in the identities
       \begin{equation}
       \begin{aligned}
       \sigma^a \bar \sigma^b &= i \bar \eta_x^{ab} \sigma^x + \delta^{ab} \mathbb{1} \\
       \bar \sigma^a \sigma^b &= i \eta_x^{ab} \sigma^x + \delta^{ab} \mathbb{1}  \;.
       \end{aligned}
     \end{equation}
       They obey the following orthogonality and self-duality relations
       \begin{equation} \label{eq:E:eta}
       \begin{aligned}
	 \bar \eta_x{}_{ab} \bar \eta_x^{cd} &=  - \epsilon_{ab}{}^{cd} +2 \delta_{[a}{}^c \delta_{b]}{}^d 
	 &\hspace{1.5cm} \bar \eta_x{}_{a b} \bar \eta_y^{ab} &= 4 \delta_{xy}  \\
	 \eta_x{}_{ab} \eta_x^{cd} &=  \epsilon_{ab}{}^{cd} +2 \delta_{[a}{}^c \delta_{b]}{}^d 
	  & \eta_x{}_{a b} \eta_y^{ab} &= 4 \delta_{xy}  \\
	  \epsilon_{xyz} \bar \eta_y{}_{ab} \bar \eta_z^{cd} &= 4 \delta_{[a}{}^{[c} \bar \eta_x{}_{b]}{}^{d]} 
	  &\epsilon_{ab}{}^{cd} \bar \eta_x{}_{cd} &= -2 \bar \eta_x{}_{ab} \\
	  \epsilon_{xyz} \eta_y{}_{ab} \eta_z^{cd} &= 4 \delta_{[a}{}^{[c} \eta_x{}_{b]}{}^{d]}
	  &\epsilon_{ab}{}^{cd} \eta_x{}_{cd} &= 2 \eta_x{}_{ab}
       \end{aligned}
     \end{equation}
       where $\epsilon_{xyz} = \epsilon_{xyz4}$ and $\epsilon_{1234}=1$.


\begin{thebibliography}{10}

\bibitem{Pestun:2007rz}
V.~Pestun, ``{Localization of gauge theory on a four-sphere and supersymmetric
  Wilson loops},'' {\em Commun.Math.Phys.} {\bf 313} (2012) 71--129,
\href{http://arXiv.org/abs/0712.2824}{{\tt 0712.2824}}.

\bibitem{Kapustin:2009kz}
A.~Kapustin, B.~Willett, and I.~Yaakov, ``{Exact Results for Wilson Loops in
  Superconformal Chern-Simons Theories with Matter},'' {\em JHEP} {\bf 1003}
  (2010) 089,
\href{http://arXiv.org/abs/0909.4559}{{\tt 0909.4559}}.

\bibitem{Jafferis:2010un}
D.~L. Jafferis, ``{The Exact Superconformal R-Symmetry Extremizes Z},'' {\em
  JHEP} {\bf 1205} (2012) 159,
\href{http://arXiv.org/abs/1012.3210}{{\tt 1012.3210}}.

\bibitem{Hama:2010av}
N.~Hama, K.~Hosomichi, and S.~Lee, ``{Notes on SUSY Gauge Theories on
  Three-Sphere},'' {\em JHEP} {\bf 1103} (2011) 127,
\href{http://arXiv.org/abs/1012.3512}{{\tt 1012.3512}}.

\bibitem{Hama:2011ea}
N.~Hama, K.~Hosomichi, and S.~Lee, ``{SUSY Gauge Theories on Squashed
  Three-Spheres},'' {\em JHEP} {\bf 1105} (2011) 014,
\href{http://arXiv.org/abs/1102.4716}{{\tt 1102.4716}}.

\bibitem{Imamura:2011wg}
Y.~Imamura and D.~Yokoyama, ``{N=2 supersymmetric theories on squashed
  three-sphere},'' {\em Phys.Rev.} {\bf D85} (2012) 025015,
\href{http://arXiv.org/abs/1109.4734}{{\tt 1109.4734}}.

\bibitem{Alday:2013lba}
  L.~F.~Alday, D.~Martelli, P.~Richmond and J.~Sparks,
  ``{Localization on Three-Manifolds},''
\href{http://arXiv.org/abs/1307.6848}{{\tt 1307.6848}}.

\bibitem{Festuccia:2011ws}
G.~Festuccia and N.~Seiberg, ``{Rigid Supersymmetric Theories in Curved
  Superspace},'' {\em JHEP} {\bf 1106} (2011) 114,
\href{http://arXiv.org/abs/1105.0689}{{\tt 1105.0689}}.

\bibitem{Klare:2012gn}
C.~Klare, A.~Tomasiello, and A.~Zaffaroni, ``{Supersymmetry on Curved Spaces
  and Holography},'' {\em JHEP} {\bf 1208} (2012) 061,
\href{http://arXiv.org/abs/1205.1062}{{\tt 1205.1062}}.

\bibitem{Dumitrescu:2012ha}
T.~T. Dumitrescu, G.~Festuccia, and N.~Seiberg, ``{Exploring Curved
  Superspace},'' {\em JHEP} {\bf 1208} (2012) 141,
\href{http://arXiv.org/abs/1205.1115}{{\tt 1205.1115}}.

\bibitem{Cassani:2012ri}
D.~Cassani, C.~Klare, D.~Martelli, A.~Tomasiello, and A.~Zaffaroni,
  ``{Supersymmetry in Lorentzian Curved Spaces and Holography},''
\href{http://arXiv.org/abs/1207.2181}{{\tt 1207.2181}}.

\bibitem{Dumitrescu:2012at}
T.~T. Dumitrescu and G.~Festuccia, ``{Exploring Curved Superspace (II)},'' {\em
  JHEP} {\bf 1301} (2013) 072,
\href{http://arXiv.org/abs/1209.5408}{{\tt 1209.5408}}.

\bibitem{Closset:2012ru}
C.~Closset, T.~T. Dumitrescu, G.~Festuccia, and Z.~Komargodski,
  ``{Supersymmetric Field Theories on Three-Manifolds},'' {\em JHEP} {\bf 1305}
  (2013) 017,
\href{http://arXiv.org/abs/1212.3388}{{\tt 1212.3388}}.

\bibitem{Hristov:2013spa}
K.~Hristov, A.~Tomasiello, and A.~Zaffaroni, ``{Supersymmetry on
  Three-dimensional Lorentzian Curved Spaces and Black Hole Holography},'' {\em
  JHEP} {\bf 1305} (2013) 057,
\href{http://arXiv.org/abs/1302.5228}{{\tt 1302.5228}}.

\bibitem{Blau:2000xg}
M.~Blau, ``{Killing spinors and SYM on curved spaces},'' {\em JHEP} {\bf 0011}
  (2000) 023,
\href{http://arXiv.org/abs/hep-th/0005098}{{\tt hep-th/0005098}}.

\bibitem{Jia:2011hw}
B.~Jia and E.~Sharpe, ``{Rigidly Supersymmetric Gauge Theories on Curved
  Superspace},'' {\em JHEP} {\bf 1204} (2012) 139,
\href{http://arXiv.org/abs/1109.5421}{{\tt 1109.5421}}.

\bibitem{Samtleben:2012gy}
H.~Samtleben and D.~Tsimpis, ``{Rigid supersymmetric theories in 4d Riemannian
  space},'' {\em JHEP} {\bf 1205} (2012) 132,
\href{http://arXiv.org/abs/1203.3420}{{\tt 1203.3420}}.

\bibitem{Liu:2012bi}
J.~T. Liu, L.~A. Pando~Zayas, and D.~Reichmann, ``{Rigid Supersymmetric
  Backgrounds of Minimal Off-Shell Supergravity},'' {\em JHEP} {\bf 1210}
  (2012) 034,
\href{http://arXiv.org/abs/1207.2785}{{\tt 1207.2785}}.

\bibitem{deMedeiros:2012sb}
P.~de~Medeiros, ``{Rigid supersymmetry, conformal coupling and twistor
  spinors},''
\href{http://arXiv.org/abs/1209.4043}{{\tt 1209.4043}}.

\bibitem{Kehagias:2012fh}
A.~Kehagias and J.~G. Russo, ``{Global Supersymmetry on Curved Spaces in
  Various Dimensions},'' {\em Nucl.Phys.} {\bf B873} (2013) 116--136,
\href{http://arXiv.org/abs/1211.1367}{{\tt 1211.1367}}.

\bibitem{Samtleben:2012ua}
H.~Samtleben, E.~Sezgin, and D.~Tsimpis, ``{Rigid 6D supersymmetry and
  localization},'' {\em JHEP} {\bf 1303} (2013) 137,
\href{http://arXiv.org/abs/1212.4706}{{\tt 1212.4706}}.

\bibitem{Kuzenko:2012vd}
S.~M. Kuzenko, ``{Symmetries of curved superspace},'' {\em JHEP} {\bf 1303}
  (2013) 024,
\href{http://arXiv.org/abs/1212.6179}{{\tt 1212.6179}}.

\bibitem{deMedeiros:2013mca}
P.~de~Medeiros and S.~Hollands, ``{Superconformal quantum field theory in
  curved spacetime},''
\href{http://arXiv.org/abs/1305.0499}{{\tt 1305.0499}}.

\bibitem{Cassani:2013dba}
D.~Cassani and D.~Martelli, ``{Supersymmetry on curved spaces and
  superconformal anomalies},''
\href{http://arXiv.org/abs/1307.6567}{{\tt 1307.6567}}.

\bibitem{Gupta:2012cy}
  R.~K.~Gupta and S.~Murthy,
  ``{All solutions of the localization equations for N=2 quantum black hole entropy},''
 {\em  JHEP} {\bf 1302} (2013) 141,
 \href{http://arXiv.org/abs/1208.6221}{{\tt 1208.6221}}.
  
  
\bibitem{Witten:1988ze}
E.~Witten, ``{Topological Quantum Field Theory},'' {\em Commun.Math.Phys.} {\bf
  117} (1988)
353.

\bibitem{Karlhede:1988ax}
A.~Karlhede and M.~Rocek, ``{Topological quantum field theory and N=2 conformal
  supergravity},'' {\em Phys.Lett.} {\bf B212} (1988)
51.

\bibitem{Balasubramanian:2000pq}
V.~Balasubramanian, E.~G. Gimon, D.~Minic, and J.~Rahmfeld, ``{Four-dimensional
  conformal supergravity from AdS space},'' {\em Phys.Rev.} {\bf D63} (2001)
  104009,
\href{http://arXiv.org/abs/hep-th/0007211}{{\tt hep-th/0007211}}.

\bibitem{Romans:1985ps}
L.~Romans, ``{Gauged N=4 supergravities in five-dimensions and their magnetovac
  backgrounds},'' {\em Nucl.Phys.} {\bf B267} (1986)
433.

\bibitem{Ohl:2010au}
T.~Ohl and C.~F. Uhlemann, ``{The Boundary Multiplet of N=4 SU(2)xU(1) Gauged
  Supergravity on Asymptotically-$AdS_5$},'' {\em JHEP} {\bf 1106} (2011) 086,
\href{http://arXiv.org/abs/1011.3533}{{\tt 1011.3533}}.

\bibitem{deWit:1979ug}
B.~de~Wit, J.~van Holten, and A.~Van~Proeyen, ``{Transformation Rules of N=2
  Supergravity Multiplets},'' {\em Nucl.Phys.} {\bf B167} (1980)
186.

\bibitem{deWit:1980tn}
B.~de~Wit, J.~van Holten, and A.~Van~Proeyen, ``{Structure of N=2
  Supergravity},'' {\em Nucl.Phys.} {\bf B184} (1981)
77.

\bibitem{LectParis}
A.~van Proeyen, ``N = 2 supergravity in d = 4, 5, 6 and its matter couplings.''
\newblock {$\newline$\tt \small
  http://itf.fys.kuleuven.be/~{}toine/LectParis.pdf}.

\bibitem{Hama:2012bg}
N.~Hama and K.~Hosomichi, ``{Seiberg-Witten Theories on Ellipsoids},'' {\em
  JHEP} {\bf 1209} (2012) 033,
\href{http://arXiv.org/abs/1206.6359}{{\tt 1206.6359}}.

\bibitem{deWit:2011gk}
B.~de~Wit and M.~van Zalk, ``{Electric and magnetic charges in N=2 conformal
  supergravity theories},'' {\em JHEP} {\bf 1110} (2011) 050,
\href{http://arXiv.org/abs/1107.3305}{{\tt 1107.3305}}.

\bibitem{Tomasiello:2011eb}
A.~Tomasiello, ``{Generalized structures of ten-dimensional supersymmetric
  solutions},'' {\em JHEP} {\bf 1203} (2012) 073,
\href{http://arXiv.org/abs/1109.2603}{{\tt 1109.2603}}.

\bibitem{Rosa:2013jja}
D.~Rosa and A.~Tomasiello, ``{Pure spinor equations to lift gauged
  supergravity},''
\href{http://arXiv.org/abs/1305.5255}{{\tt 1305.5255}}.

\bibitem{kirilrota}
K.~Hristov and A.~Rota, ``to appear,''.

\bibitem{Alday:2009aq}
L.~F. Alday, D.~Gaiotto, and Y.~Tachikawa, ``{Liouville Correlation Functions
  from Four-dimensional Gauge Theories},'' {\em Lett.Math.Phys.} {\bf 91}
  (2010) 167--197,
\href{http://arXiv.org/abs/0906.3219}{{\tt 0906.3219}}.

\bibitem{Hama:2013ama}
N.~Hama and K.~Hosomichi, ``{AGT relation in the light asymptotic limit},''
\href{http://arXiv.org/abs/1307.8174}{{\tt 1307.8174}}.

\bibitem{Nekrasov:2003rj}
N.~Nekrasov and A.~Okounkov, ``{Seiberg-Witten theory and random partitions},''
\href{http://arXiv.org/abs/hep-th/0306238}{{\tt hep-th/0306238}}.

\bibitem{Nekrasov:2002qd}
N.~A. Nekrasov, ``{Seiberg-Witten prepotential from instanton counting},'' {\em
  Adv.Theor.Math.Phys.} {\bf 7} (2004) 831--864,
\href{http://arXiv.org/abs/hep-th/0206161}{{\tt hep-th/0206161}}.

\end{thebibliography}

\providecommand{\href}[2]{#2}

\end{document}